\documentclass[twocolumn,amsmath,amssymb,showpacs,pre,superscriptaddress]
{revtex4-1}
\usepackage[utf8]{inputenc}
\usepackage{color,graphicx,subfigure,t1enc,grffile}

\newcommand{\eqnref}[1]{Eq.~\ref{eq:#1}}
\newcommand{\picref}[1]{Fig.~\ref{fig:#1}}
\newcommand{\tabref}[1]{Tab.~\ref{tab:#1}}
\newcommand{\chref}[1]{\ref{ch:#1}}

\bibstyle{apsrev}

\begin{document}

\title[short]{Accurate lubrication corrections for spherical and
  non-spherical particles in discretized fluid simulations}

\author{F. Janoschek}
\email{f.j@noschek.de}
\affiliation{Department of Applied Physics, Eindhoven University of
  Technology, P.\,O. Box 513, 5600\,MB Eindhoven, The Netherlands}
\author{J. Harting}
\email{j.harting@tue.nl}
\affiliation{Department of Applied Physics, Eindhoven University of
  Technology, P.\,O. Box 513, 5600\,MB Eindhoven, The Netherlands}
\affiliation{Institute for Computational Physics, University of Stuttgart,
  Allmandring 3, 70569 Stuttgart, Germany}
\author{F. Toschi}
\email{f.toschi@tue.nl}
\affiliation{Department of Applied Physics, Eindhoven University of
  Technology, P.\,O. Box 513, 5600\,MB Eindhoven, The Netherlands}
\affiliation{CNR-IAC, Via dei Taurini 19, 00185 Rome, Italy}

\date{\today}

\begin{abstract}
  Discretized fluid solvers coupled to a Newtonian dynamics method are
  a popular tool to study suspension flow. As any simulation technique
  with finite resolution, the lattice Boltzmann method, when coupled
  to discrete particles using the momentum exchange method, resolves
  the diverging lubrication interactions between surfaces near contact
  only insufficiently. For spheres, it is common practice to account
  for surface-normal lubrication forces by means of an explicit
  correction term. A method that additionally covers all further
  singular interactions for spheres is present in the literature as
  well as a link-based approach that allows for more general shapes
  but does not capture non-normal interactions correctly.
  In this paper, lattice-independent lubrication corrections for
  aspherical particles are outlined, taking into account all leading
  divergent interaction terms. An efficient implementation for
  arbitrary spheroids is presented and compared to purely normal and
  link-based models. Good consistency with Stokesian dynamics
  simulations of spheres is found. The non-normal interactions affect
  the viscosity of suspensions of spheres at volume fractions
  $\Phi\ge0.3$ but already at $\Phi\ge0.2$ for spheroids. Regarding
  shear-induced diffusion of spheres, a distinct effect is found at
  $0.1\le\Phi\le0.5$ and even increasing the resolution of the radius
  to $8$ lattice units is no substitute for an accurate modeling of
  non-normal interactions.
\end{abstract}

\maketitle

\section{Introduction}

The dynamics of particles suspended in a fluid plays an important role
for a large set of problems ranging from sedimentation and
fluidization processes in industrial-scale chemical reactors to
capillary blood flow in human microcirculation. Especially the
shear-induced mass transport in suspensions of non-spherical
particles, such as blood cells, has grown to a very active field of
research recently~\cite{lopez07, clausen11, kruger12, zhao12,
  grandchamp13, omori13}. All these examples have in common that the
gap between the surfaces of either two particles or between one
particle and the geometry confining the flow frequently becomes small
as compared to the particle size. The consequence are hydrodynamic
short-range interactions mediated by the interstitial fluid that
increase in strength as the distance of the surfaces decreases and
that can play an important role in suspension rheology~\cite{yeo13}
but also in the dynamics of the suspended particles
themselves~\cite{metzger13}. The smallness of the gap between the
surfaces allows for the assumption of Stokes flow. Thus, the forces
and torques on the surfaces appear as linear functions of their
translation and rotation velocities which is most conveniently
formulated in terms of a resistance matrix. Furthermore, the smallness
of the gap allows a lubrication-theoretical treatment of the
interactions. While the lubrication limit is treated already by
Goldman \textit{et al.}~\cite{goldman67} in the case of a sphere next
to a plane wall, the work by Cox~\cite{cox74} is the first to consider
arbitrary yet smooth and convex surfaces. Approximating the surfaces
at their points of closest approach as polynomials of second order,
Cox~\cite{cox74} studies the divergence behavior of the resistance
matrix for vanishing gap widths $h$ and presents explicit expressions
for the leading-order terms of most of the matrix elements. It is
found that while the surface-normal force induced by a relative
translation of the surfaces along the same direction diverges as
$h^{-1}$, all other interactions show a weaker divergence proportional
to $\ln h$ or even remain finite. Claeys and Brady~\cite{claeys89}
complete the study by Cox~\cite{cox74}, taking into account the third-
and fourth-order expansion coefficients of the local surface geometry
which are required to compute all diverging terms for all matrix
elements. The results are employed later for the Stokesian dynamics
simulation of suspensions of prolate spheroids by the same
authors~\cite{claeys93}. A computationally more efficient model for
oblate spheroids that neglects long-range hydrodynamic interactions
and for the computation of lubrication interactions locally
approximates the interacting surfaces as spheres is proposed by
Bertevas \textit{et al.}~\cite{bertevas10}.

Stokesian dynamics simulations are restricted to the creeping flow
regime. To model suspension flow at finite Reynolds numbers, the
lattice Boltzmann (LB) method~\cite{succi01}, especially when used in
connection with the momentum exchange method originating from
Ladd~\cite{ladd94,ladd94b}, has emerged as an increasingly popular
technique during the last two decades that further allows for a
comparably easy parallel implementation and for complex boundary
conditions~\cite{ladd01,aidun10}. Since the method describes the fluid
only at discrete nodes of a lattice with finite spatial resolution it
cannot account for lubrication interactions at arbitrarily small
particle separations directly. Already resolving them at separations
of $\sim10\,\%$ of the particle radius would require lattice
resolutions that are finer and computationally more expensive than the
ones necessary to obtain accurate drag coefficients and particle
interactions at larger separations~\cite{ladd94b, ladd01, kunert10,
  kunert11}. Similar problems arise also in other simulation methods
with finite resolution, such as finite element
methods~\cite{schwarzer95}, stochastic rotation
dynamics~\cite{hecht05}, or dissipative particle
dynamics~\cite{martys05}. In the case of spherical particles it is
common practice to address these issues by correcting the LB method
for particles near contact with the asymptotic expressions known from
lubrication theory. While many implementations correct only for normal
lubrication forces resulting from a central approach of the
spheres~\cite{ladd97,hyvaluoma05,kromkamp06}, Nguyen and
Ladd~\cite{nguyen02} account for the leading divergence terms of the
weaker non-normal interactions as well. To the best of the authors'
knowledge, a comparably accurate method for aspherical particles does
not exist up to now. In fact, present applications of the LB and
momentum exchange method to suspensions of aspherical particles often
do not account for lubrication interactions explicitly~\cite{qi02},
ignore the torque resulting from asymmetric
encounters~\cite{gunther13}, or defer the description of short-range
interactions to an empirical model~\cite{janoschek10}. On the other
hand, Ding and Aidun~\cite{ding03} introduce a method for lubrication
correction that is based on the interconnecting lattice links between
particles near contact and thus is directly applicable to aspherical
particles. The method is employed later in simulations of deformable
particles~\cite{macmeccan09}. More recently, however, drawbacks of the
link-based approach are stated to be the demand for a relatively large
minimum lattice resolution~\cite{aidun10} and the misestimation of
non-normal lubrication interactions~\cite{clausen10b}.

It therefore appears that the present literature shows some
uncertainty regarding the degree of accuracy actually required from
lubrication corrections in LB simulations as well as regarding how to
implement a sufficiently accurate lubrication model for aspherical
particles. The goal of this work is to mitigate these
uncertainties. In section~\chref{methods} below the LB method and the
momentum exchange method are briefly introduced, followed by an
outline of contact-based lubrication corrections for spheres and of a
link-based lubrication model. In section~\chref{lub} the
implementation of accurate lubrication corrections for aspherical
particles following the analytical work by Cox~\cite{cox74} and by
Claeys and Brady~\cite{claeys89} is demonstrated for the case of
spheroids. Section~\chref{results} compares the different lubrication
models with respect to the accuracy of two-particle interactions and
with respect to the shear-induced diffusion and the viscosity of a
suspension as examples for one observable that examines the dynamics
of single particles and one averaged observable, all with the focus on
the effect of non-normal lubrication corrections. Conclusions are
drawn in section~\chref{conclusions}. The appendix provides a
compilation of the diverging terms in the resistance matrix that are
given already by Cox~\cite{cox74} and the remaining leading terms
first computed by Claeys and Brady~\cite{claeys89}.

\section{Lattice Boltzmann method for suspensions of solid
  particles}\label{ch:methods}

Historically, the LB method originates from lattice gas cellular
automata. A comprehensive introduction is available in the book by
Succi~\cite{succi01}. Time $t$ is discretized in steps $\delta t=1$,
space in positions $\mathbf{x}$ on a regular lattice defined by a
finite set of $q$ discrete velocity vectors $\mathbf{c}_r$ with
$r=1,\ldots,q$. Fluid particles at position $\mathbf{x}$ and time $t$
traveling along $\mathbf{c}_r$ are represented by the discretized
single-particle distribution function $n_r(\mathbf{x},t)$. The
algorithm to propagate $n_r(\mathbf{x},t)$ in time prescribes the
repeated consecutive execution of the advection step
\begin{equation}
  \label{eq:methods:advection}
  n_r(\mathbf{x}+\mathbf{c}_r,t+\delta t)
  =
  n^*_r(\mathbf{x},t)
\end{equation}
and the collision step
\begin{equation}
  \label{eq:methods:collision}
  n^*_r(\mathbf{x},t)
  =
  n_r(\mathbf{x},t)
  -
  \Omega
  \text{ ,}
\end{equation}
the latter producing the post-collision distribution
$n^*_r(\mathbf{x},t)$. \eqnref{methods:collision} and
\eqnref{methods:advection} together form the LB equation. For the sake
of simplicity, the Bhatnagar-Gross-Krook collision term
\begin{equation}\label{eq:methods:lbgk}
  \Omega
  =
  \frac
  {n_r(\mathbf{x},t)-
    n_r^\mathrm{eq}(\rho(\mathbf{x},t),\mathbf{u}(\mathbf{x},t))}
  {\tau}
\end{equation}
with a single-relaxation time $\tau$ is employed. It relies on a
second-order expansion of the Maxwell-Boltzmann equilibrium
distribution
\begin{equation}\label{eq:methods:equilibrium}
  n_r^\mathrm{eq}(\rho,\mathbf{u})
  =
  \rho\alpha_{c_r}
  \left[
  1
  +\frac{\mathbf{c}_r\mathbf{u}}{c_\text{s}^2}
  +\frac{\left(\mathbf{c}_r\mathbf{u}\right)^2}{2c_\text{s}^4}
  -\frac{\mathbf{u}^2}{2c_\text{s}^2}
  \right]
\end{equation}
with a speed of sound $c_\mathrm{s}$. The local density
\begin{equation}
  \rho(\mathbf{x},t)=\sum_rn_r(\mathbf{x},t)
\end{equation}
and velocity
\begin{equation}
  \mathbf{u}(\mathbf{x},t)=
  \frac{\sum_rn_r(\mathbf{x},t)\mathbf{c}_r}{\rho(\mathbf{x},t)}
\end{equation}
are calculated as moments of the fluid distribution. In the following,
the three-dimensional D3Q19 lattice~\cite{qian92} is applied for which
$q=19$ and the lattice weights
\begin{equation}
  \alpha_{c_r}
  =
  \left\{
    \begin{array}{l@{\quad\text{for }c_r=\,\,}l}
      1/3 & 0\\
      1/18 & 1\\
      1/36 & \sqrt{2}
    \end{array}
  \right.
  \text{ .}
\end{equation}
In a Chapman-Enskog expansion it can be shown that
$\mathbf{u}(\mathbf{x},t)$ as obtained from the method follows the
incompressible Navier-Stokes equations with a kinematic viscosity
$\nu=\left(\tau-\frac{1}{2}\right)c_\mathrm{s}^2$ in the limit of
small Mach numbers $\mathrm{Ma}=u/c_\mathrm{s}$ with
$c_\mathrm{s}=1/\sqrt{3}$~\cite{qian92}.

In the momentum exchange method~\cite{ladd01}, particle volumes of in
principle arbitrary shape are discretized on the lattice as outlined
in \picref{methods:sketch}(a) and coupled to the fluid via the links
crossing the resulting particle-fluid interface. While the original
work~\cite{ladd94,ladd94b} treats particles as fluid-filled shells,
the inner fluid is removed in later
implementations~\cite{aidun98,nguyen02} including the one described
here. A no-slip boundary moving with velocity $\mathbf{v}_\mathrm{b}$
is established by a mid-link bounce-back rule
\begin{equation}\label{eq:methods:moving-bounce-back}
  n_r(\mathbf{x}+\mathbf{c}_r,t+\delta t)
  =
  n^*_{\bar{r}}(\mathbf{x}+\mathbf{c}_r,t)
  +
  C
\end{equation}
with a first-order velocity correction~\cite{ladd94}
\begin{equation}\label{eq:methods:bounce-back-correction}
  C
  =
  \frac
  {2\alpha_{c_r}}
  {c_\text{s}^2}
  \rho(\mathbf{x}+\mathbf{c}_r,t)
  \,
  \mathbf{c}_r\mathbf{v}_\mathrm{b}
\end{equation}
that replaces \eqnref{methods:advection} where required to prevent
advection out of a particle site $\mathbf{x}$. The index $\bar{r}$ is
defined by
$\mathbf{c}_{\bar{r}}\equiv-\mathbf{c}_r$. \eqnref{methods:moving-bounce-back}
with \eqnref{methods:bounce-back-correction} is easily shown to be
consistent with \eqnref{methods:equilibrium}. The reduction of fluid
momentum by each bounce-back process
\begin{equation}\label{eq:methods:momentum-exchange}
  \mathrm{d}\mathbf{p}
  =
  \left(
    2n_{\bar{r}}
    +
    C
  \right)
  \mathbf{c}_{\bar{r}}
  \text{ ,}
\end{equation}
is transferred to the respective particle. According to the choice of
unit time steps it is equal to the resulting
force. \eqnref{methods:momentum-exchange} can be seen as a discretized
traction vector and is therefore used to compute the hydrodynamic
force and torque on the particle. When, due to particle motion, new
fluid sites are covered, the fluid at those sites is deleted. When a
site formerly occupied by a particle is freed, new fluid is created
according to \eqnref{methods:equilibrium} with an average fluid
density $\rho$ and $\mathbf{u}$ estimated according to the rigid-body
motion of the particle. Momentum conservation is ensured
instantaneously by an appropriate force on the particle. The data
published by Ladd~\cite{ladd94b} for an input radius $R=4.5$ defining
the discretization on the lattice suggests that the effective
hydrodynamic radius $R^*$ defined via the drag coefficient deviates
least from $R$ for a relaxation time somewhat below $\tau=1$. In the
following, $\tau=1$ is chosen and no effort is made to re-calibrate
particle radii using $R^*$. For spherical particles both the relative
deviations of the translational drag coefficient from the expected
value and its fluctuations due to the aforementioned discretization
changes are found to be below $10\,\%$ already at a resolution of the
sphere radius of only $R=2.5$ lattice
sites~\cite{ladd94b,janoschek11b}.

Problems arise when particle surfaces approach closely. The
short-range interactions are truthfully described down to a gap width
of only about $1$ lattice spacing~\cite{ladd94b}. At shorter
distances, the expected divergence is not reproduced, instead the
friction coefficients stay approximately constant to the value
achieved at a distance of $1$~\cite{ladd94b}. The observation can be
understood as depicted in \picref{methods:sketch}(a): a gap width
between both surfaces of about $1$ lattice spacing is the distance
below which direct links between both particles emerge. Further
approach does not lead to changes in the fluid site configuration in
the gap which could cause an increase in the interaction forces.

\begin{figure}
  \includegraphics[width=0.7\columnwidth]{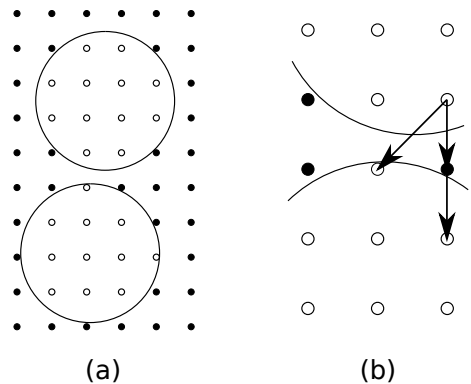}
  \caption{\label{fig:methods:sketch}(a) The discretized
    representation on the lattice leads to unresolved short-range
    hydrodynamic interactions between particles near contact. In the
    link-wise approach by Ding and Aidun~\cite{ding03}, correction
    forces are computed for single links and for pairs of identical
    links connecting the particles as visualized in (b). For clarity,
    not all interconnecting links are drawn.}
\end{figure}

Ladd addresses the issue later~\cite{ladd97,ladd01} by employing the
dominating divergence term $\sim h^{-1}$ of the normal force induced
by the central approach of two spheres at a gap distance $h=r_{ij}-2R$
as a correction
\begin{equation}\label{eq:methods:llub}
  \mathbf{f}_{ij}=-\mathbf{f}_{ji}
  =
  -
  \frac{3\pi\mu R^2}{2}
  \hat{\mathbf{r}}_{ij}
  \hat{\mathbf{r}}_{ij}
  \cdot
  (\mathbf{v}_i-\mathbf{v}_j)
  \left[
    \frac{1}{h}
    -
    \frac{1}{\Delta_\mathrm{c}}
  \right]
\end{equation}
to the hydrodynamic force on sphere $i$ due to another sphere $j$ for
which $h<\Delta_\mathrm{c}$. The center displacement is
$\mathbf{r}_{ij}=\mathbf{r}_i-\mathbf{r}_j$, the related unit vector
$\hat{\mathbf{r}}_{ij}=\mathbf{r}_{ij}/r_{ij}$, $\mathbf{v}_i$ and
$\mathbf{v}_j$ denote the particles' translational velocities, and
$\mu=\rho\nu$ refers to the dynamic viscosity of the suspending
medium. As a cut-off parameter, $\Delta_\mathrm{c}$ represents the
separation below which the LB method alone does not sufficiently cover
hydrodynamic interactions anymore. $\Delta_\mathrm{c}$ is a function
of $\tau$~\cite{nguyen02}.  For $\tau=1$, a value
$\Delta_\mathrm{c}=2/3$ is
suggested~\cite{nguyen02,ladd01}. Corrections equivalent to
\eqnref{methods:llub} for spheres of possibly differing radii are
introduced also for the forces and torques resulting from rotation and
non-normal translation by Nguyen and Ladd~\cite{nguyen02}. These,
however, diverge only as $\ln h$ and require separate cut-off
parameters~\cite{nguyen02}.

An implicit integration scheme for the particle trajectories is one
way to maintain numerical stability in the presence of clusters of
particles with strong lubrication
interactions~\cite{ladd01,nguyen02}. In the implementation employed
here, the time step for the particle update is decoupled from the LB
time step instead which allows its reduction, to typically $1/10$:
while the forces due to the momentum exchanged with the LB fluid via
\eqnref{methods:momentum-exchange} remain constant over one LB step,
the particle positions and velocities are updated in accordance with
the strongly varying explicit lubrication forces at a finer temporal
resolution.

\label{ch:methods:link-based}Ding and Aidun~\cite{ding03} propose an
alternative method for lubrication corrections that is based on the
interconnecting lattice links between particles and therefore does not
require analytical knowledge of the particles' asymptotic resistance
functions~\cite{aidun10}. In this method, a partial lubrication force
\begin{equation}\label{eq:methods:dlub}
  \mathrm{d}\mathbf{f}_{ij}=-\mathrm{d}\mathbf{f}_{ji}
  =
  -
  \frac
  {3\bar{q}\mu}
  {2c_r^2\lambda^*}
  \hat{\mathbf{c}}_r
  \hat{\mathbf{c}}_r
  \cdot
  (\mathbf{v}_{\mathrm{b}i}-\mathbf{v}_{\mathrm{b}j})
  \left[
    \frac{1}{h^{*2}}
    \!-\!
    \frac{1}{c_r^2}
  \right]
\end{equation}
with $\hat{\mathbf{c}}_r=\mathbf{c}_r/c_r$ is applied locally to
particle $i$ for all links $\mathbf{c}_r$ that end on a site belonging
to $i$ and stem from either a site of particle $j$ or a fluid site at
the center of two links $2\mathbf{c}_r$ originating from $j$. Both
possibilities are outlined in \picref{methods:sketch}(b). The gap
distance $h^*$ is the distance between the intersections of both
theoretical particle surfaces with the lattice link or the pair of
links and therefore is typically larger than the actual minimum
gap. The correction is applied only where $h^*<c_r$. The velocities
$\mathbf{v}_{\mathrm{b}i}$ and $\mathbf{v}_{\mathrm{b}j}$ at the
intersection points are computed from the particles' rigid body
motion. Different from \eqnref{methods:llub}, the model can produce
non-central forces depending on the link direction $\mathbf{c}_r$. The
curvature $\lambda^*$ is obtained as the mean of both surfaces and
$\bar{q}=0.6$ is an empiric weighting factor~\cite{ding03}. Despite
its generality, the model is applied and validated initially only in
the case of normal approach of spheres and cylinders towards each
other and towards a flat wall~\cite{ding03}. For centrally approaching
spheres, analytical consistency with the known asymptotic behavior is
demonstrated~\cite{ding03}.

\eqnref{methods:dlub} is applied to deformable particles
later~\cite{macmeccan09} but Clausen~\cite{clausen10b} notes an
erroneous divergence of tangential interactions as $h^{-1}$ which he
fixes, along with further modifications, by the introduction of an
average normal direction $\mathbf{n}_\mathrm{avg}$ of the surfaces.
Both the difference of the boundary velocities and the link-wise gap
are projected along $\mathbf{n}_\mathrm{avg}$ and the resulting force
is applied in the same direction. Thereby, however, locally tangential
lubrication corrections are removed. Moreover, the discretization of
lubrication interactions onto interconnecting lattice links that
appear and vanish as particles move is found to cause
instabilities~\cite{clausen10b} that apparently prevent usage of the
link-wise model in at least some of the subsequent
work~\cite{clausen11}.

\section{Contact-based lubrication corrections for
  spheroids}\label{ch:lub}

Though many of the following ideas could be applied to particles of
other convex shapes as well, spheroids with half axes $R_\parallel$
and $R_\perp$ parallel and perpendicular to their axis of rotational
symmetry will be treated here. A convenient parametrization of the
surface is
\begin{equation}\label{eq:lub:spheroid}
  \tilde{\mathbf{y}}(p,q)
  =
  \left(
    \begin{array}{c}
      R_\perp\cos p\cos q\\
      R_\perp\cos p\sin q\\
      R_\parallel\sin p
    \end{array}
  \right)
  \text{ ,}
\end{equation}
where the tilde indicates a representation in the body-fixed reference
frame where the origin is at the center of the particle and the
$\tilde{y}_3$-direction oriented along its axis of rotational
symmetry. For symmetry reasons, the directions of principal curvature
are the tangential directions
\begin{equation}\label{eq:lub:curvature-directions}
  \tilde{\mathbf{s}}_1(p,q)=\partial_p\tilde{\mathbf{y}}(p,q)
  \text{\quad and\quad}
  \tilde{\mathbf{s}}_2(p,q)=\partial_q\tilde{\mathbf{y}}(p,q)
  \text{ .}
\end{equation}
The respective radii of curvature
\begin{equation}\label{eq:lub:curvature-radii}
  S_1(p) = \frac{s_1(p)^3}{R_\parallel R_\perp}
  \text{\quad and\quad}
  S_2(p) = \frac{R_\perp s_1(p)}{R_\parallel}
  \text{ ,}
\end{equation}
with
\begin{equation}
  s_1(p)=|\tilde{\mathbf{s}}_1(p)|
  =
  \sqrt{R_\perp^2\sin^2p + R_\parallel^2\cos^2p}
\end{equation}
can be obtained from a second-order expansion of
\eqnref{lub:spheroid}.

\subsection{Minimum gap between two spheroids}

The strong dependence of lubrication interactions on the minimum
separation between two particles $i$ and $j$ requires precise
knowledge of the magnitude and direction of the minimum gap vector
$\mathbf{h}$. For two spheres with radii $R_i$ and $R_j$, the result
trivially is $\mathbf{h}=-(r_{ij}-R_i-R_j)\hat{\mathbf{r}}_{ij}$,
where the minus sign ensures a direction away from particle
$i$. Already for spheroids the problem in general is considerably more
intricate. One solution is to follow the iterative procedure presented
by Lin and Han~\cite{lin02} for the distance between two
ellipsoids. As illustrated in \picref{lub:lin02}, the method involves
the re-positioning of tangent spheres along the inner surface of each
ellipsoid to minimize the distance of the sphere centers and thus the
gap between the ellipsoids. A sufficient requirement for convergence
of the method is that each sphere is completely contained in the
respective ellipsoid~\cite{lin02}. This is achieved easily if the
radius is chosen as the minimum radius of curvature
\begin{equation}\label{eq:lub:minimum-radius}
  \bar{S}
  =
  \left\{
    \begin{array}[c]{l@{\quad\text{for }}l}
      R_\parallel^2/R_\perp & R_\parallel<R_\perp\\
      R & R_\parallel=R_\perp=R\\
      R_\perp^2/R_\parallel & R_\parallel>R_\perp
    \end{array}
  \right.
  \text{ .}
\end{equation}
Initially, the spheres are placed tangent in the intersection points
of the spheroid surfaces with the line connecting the spheroid
centers. In the iteration, the tangent points $\mathbf{y}_i$ and
$\mathbf{y}_j$ are repeatedly updated to become the intersections of
the spheroid surfaces with the line connecting the current centers
$\mathbf{z}_i$ and $\mathbf{z}_j$ of the spheres. The iteration stops
once the angles $\theta_i$ and $\theta_j$ between the outward pointing
surface unit normals $\hat{\mathbf{n}}_i$ in $\mathbf{y}_i$ and
$\hat{\mathbf{n}}_j$ in $\mathbf{y}_j$ and the vector
$\mathbf{z}_{ij}=\mathbf{z}_i-\mathbf{z}_j$ between the sphere centers
approximate zero. The converged surface positions of minimum distance
are referred to as $\mathbf{y}_i^*$ and $\mathbf{y}_j^*$ in the
following. Practically, the convergence criterion is implemented as
the requirement that both
\begin{equation}\label{eq:lub:convergence}
  \cos\theta_i=-\hat{\mathbf{n}}_i\cdot\hat{\mathbf{z}}_{ij}>1-\epsilon
  \text{ ,}
\end{equation}
with $\hat{\mathbf{z}}_{ij}=\mathbf{z}_{ij}/z_{ij}$ and
\eqnref{lub:convergence} with indices $i$ and $j$ swapped hold. Due to
the small magnitude of $\partial\cos\theta_i/\partial\theta_i$ near
$\theta_i=0$, a rather low value of $\epsilon=10^{-6}$ is chosen to
ensure accuracy. The resulting gap vector is
$\mathbf{h}=-(z_{ij}-2\bar{S})\hat{\mathbf{z}}_{ij}$.

For overlapping particles, Lin and Han~\cite{lin02} effectively state
that $h=0$ and abstain from a further treatment. When implemented
accordingly, their method as described above is, however, stable also
for configurations were particles overlap to an amount small as
compared to their radii of curvature. Then, $\mathbf{h}$ still points
from the bulk of particle $i$ to $j$ but the scalar gap
$z_{ij}-2\bar{S}$ is negative. This is an important feature that could
be used to model elastic contact forces but also to enhance the
stability of dense simulations of infinitely stiff particles with a
finite time step where minimal amounts of overlap are surely undesired
but sometimes inevitable.

\eqnref{lub:minimum-radius} ensures that the tangent spheres assumed
in the algorithm lie within the respective spheroids which is noted to
be sufficient for convergence but not necessarily optimal with respect
to the number of iterations required~\cite{lin02}. In many cases
typical for the spheroids simulated here, convergence can indeed be
speeded up dramatically by adjusting $\bar{S}$ for each surface to the
smaller of the two local principal radii of curvature in
\eqnref{lub:curvature-radii}. Thus, a two-fold strategy is being
followed: first, up to $N_1$ iterations with
$\bar{S}=\min\{S_1(p),S_2(p)\}$ are performed. Since due to the
potentially enlarged $\bar{S}$ and the allowance for overlap a
meaningful result is not guaranteed, two consistency checks are
performed whenever \eqnref{lub:convergence} indicates convergence: it
is required that $\hat{\mathbf{n}}_j\cdot\mathbf{r}_{ij}>0$ and
$\hat{\mathbf{n}}_i\cdot\mathbf{r}_{ij}<0$ so the outward-directed
surface normals point towards the center of the other particle. In
case of overlap, $\mathbf{y}_i^*$ has to be contained within particle
$j$ and vice versa. If at least one of the conditions fails or
convergence is not achieved yet, the algorithm restarts with the more
conservative $\bar{S}$ from \eqnref{lub:minimum-radius} for a maximum
of $N_2$ iterations. Here, $N_1=10$ and $N_2=1000$ is chosen as a set
of values that for particles with aspect ratio
$\Lambda=R_\parallel/R_\perp=1/3$ results in an average total number
of iterations per gap computation of almost $40$. This is a reduction
of about $30\,\%$ as compared to only one iteration with fixed radii
defined by \eqnref{lub:minimum-radius}. So far, no configuration
without significant overlap is known where the combined procedure with
these parameters fails to converge.

With a cut-off gap such as $\Delta_\mathrm{c}$ in
\eqnref{methods:llub}, it is necessary to determine $\mathbf{h}$ once
during every (particle-)time step for every pair of particles with a
minimum separation that is potentially smaller than
$\Delta_\mathrm{c}$. All pairs with a center distance
$r_{ij}>2\max\{R_\parallel,R_\perp\}+\Delta_\mathrm{c}$ can clearly be
excluded. A computation is unnecessary also if
$r_{ij}>\max\{R_\parallel,R_\perp\}+\Delta_\mathrm{c}\equiv
R_\mathrm{c}$ but at the same time particle $i$ has no intersection
with a plane normal to $\mathbf{r}_{ij}$ in a distance of
$R_\mathrm{c}$ away from $\mathbf{r}_j$ in the direction of $i$ or
vice versa. This can be tested in a comparably inexpensive way and
avoids unneeded computations of $\mathbf{h}$, especially in the case
of highly aspherical particles. A further very effective approach for
optimization consists in starting the iteration to obtain $\mathbf{h}$
not from the particle centers as explained above but using the
converged tangent positions $\mathbf{y}_i^*$ and $\mathbf{y}_j^*$ of
the previous computation instead. Since particle configurations do not
change much during one time step this often enables convergence within
only one iteration and results in a further reduction of the average
total number of iterations per gap computation from almost $40$ to
less than $1.01$ for $\Lambda=1/3$. The same result is found in
typical simulations of particles with other aspect ratios
$1/4\le\Lambda\le4$.

\begin{figure}
  \includegraphics[width=0.95\columnwidth]{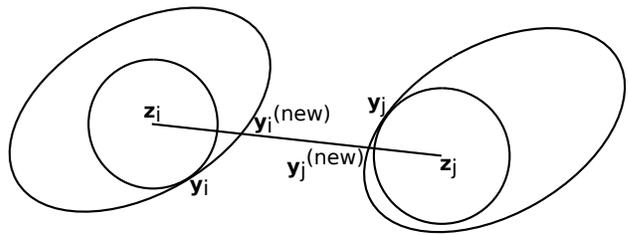}
  \caption{\label{fig:lub:lin02}Two-dimensional outline of the
    iterative approach described by Lin and Han~\cite{lin02} to find
    the minimum gap between the surfaces of two ellipsoids. (idea for
    figure from Ref.~\cite{lin02})}
\end{figure}

\subsection{Diverging lubrication interactions}

Cox~\cite{cox74} and later Claeys and Brady~\cite{claeys89} expand the
surfaces of two particles $i$ and $j$ near contact around their
closest points $\mathbf{y}_i^*$ and $\mathbf{y}_j^*$ as
\begin{equation}\label{eq:lub:surface-i}
  x_3
  =
  -\frac{x_1^2}{2S_1}
  -\frac{x_2^2}{2S_2}
  -\sum_{k=0}^3\Gamma_kx_1^{3-k}x_2^k
  +\mathcal{O}(r^4)
\end{equation}
and
\begin{equation}\label{eq:lub:surface-j}
  x_3'
  =
  \frac{x_1'^2}{2S_1'}
  +\frac{x_2'^2}{2S_2'}
  +\sum_{k=0}^3\Gamma_k'x_1'^{3-k}x_2'^k
  +\mathcal{O}(r'^4)
\end{equation}
in the tangential coordinates $x_1$ and $x_2$ that are measured along
the directions of principal curvature of the respective particle,
given in \eqnref{lub:curvature-directions}. Here and in the following,
primed variables refer to the surface of particle $j$ and unprimed
ones to the one of $i$. The $\Gamma_{0-3}$ are the coefficients of the
third-order terms in the expansion. The error terms scale as the
fourth power of the tangential distance $r=\sqrt{x_1^2+x_2^2}$ from
the gap position. In fact, Cox~\cite{cox74} in general considers only
the quadratic terms in \eqnref{lub:surface-i} and
\eqnref{lub:surface-j} while Claeys and Brady~\cite{claeys89} expand
up to even fourth-order. Here, only second and third order are taken
into account. The fourth-order coefficients are needed only for a
weakly diverging $\ln h$ contribution to surface-normal lubrication
forces~\cite{claeys89} which at small gaps $h$ must be dominated by
the $h^{-1}$ term (see also appendix). As is shown in
\picref{lub:cox74}, $\phi$ denotes the angle between the principal
directions of curvature of both surfaces and both $x_3$-axes point in
the direction of $\mathbf{h}$ while the origins, $\mathbf{y}_i^*$ and
$\mathbf{y}_j^*$, obviously differ.

\begin{figure}
  \includegraphics[width=0.95\columnwidth,trim=0cm 0.5cm 0cm 2.5cm,clip]
  {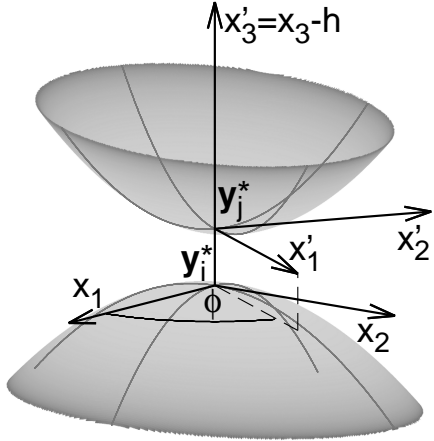}
  \caption{\label{fig:lub:cox74}The coordinate axes as defined by
    Cox~\cite{cox74} and by Claeys and Brady~\cite{claeys89} for the
    approximation of the two surfaces at the minimum gap as bivariate
    cubic polynomials. (idea for figure from Ref.~\cite{claeys89})}
\end{figure}

In Ref.~\cite{cox74} and Ref.~\cite{claeys89}, the asymptotic behavior
for small $h$ of the force $\bar{\mathbf{F}}$ and torque
$\bar{\mathbf{T}}$ on each of the surfaces $i$ and $j$ is analyzed in
dependence of the translational and rotational velocities
$\bar{\mathbf{V}}$ and $\bar{\boldsymbol{\Omega}}$ of both $i$ and
$j$. The bar on vector quantities is used to indicate a representation
in the local surface coordinates $\{x_1,x_2,x_3\}$ associated with
particle $i$. Since the diverging terms depend only on the relative
velocities, the relation can be expressed as
\begin{equation}\label{eq:lub:resistance}
  \left(
    \begin{array}{c}
      \bar{\mathbf{F}}_i\\
      \bar{\mathbf{T}}_i
    \end{array}
  \right)
  =
  -
  \left(
    \begin{array}{c}
      \bar{\mathbf{F}}_j\\
      \bar{\mathbf{T}}_j
    \end{array}
  \right)
  =
  \mu\mathbf{K}
  \left(
    \begin{array}{c}
      \bar{\mathbf{V}}_j-\bar{\mathbf{V}}_i\\
      \bar{\boldsymbol{\Omega}}_j-\bar{\boldsymbol{\Omega}}_i
    \end{array}
  \right)
  +
  \mathcal{O}(h^0)
  \text{ ,}
\end{equation}
with a $6\times6$ resistance matrix
$\mathbf{K}$~\cite{cox74,claeys89}. $\mathbf{K}$ is symmetric and some
of the elements are zero as the corresponding forces and torques
remain finite upon contact. Thus, there are only $16$ independent
non-zero matrix elements $K_{\gamma\delta}$ which are functions of the
gap $h$, the angle $\phi$, and the expansion coefficients in
\eqnref{lub:surface-i} and \eqnref{lub:surface-j}. These
$K_{\gamma\delta}$ can be obtained from the
literature~\cite{cox74,claeys89} but are listed in the appendix for
the sake of completeness. Since a part of the lubrication interactions
is already accounted for by the LB method, only a correction for small
gaps is required. Analogously to Ladd's initial approach in
\eqnref{methods:llub}~\cite{ladd97,nguyen02}, a corrective resistance
matrix $\tilde{\mathbf{K}}$ is constructed from the differences
\begin{equation}\label{eq:lub:resistance-correction}
  \tilde{K}_{\gamma\delta}(h)
  =
  \left\{
    \begin{array}{l@{\quad\text{for }}l}
      K_{\gamma\delta}(h)
      -
      K_{\gamma\delta}(\Delta_{\gamma\delta})
      &
      h<\Delta_{\gamma\delta}\\
      0 & h\ge\Delta_{\gamma\delta}
  \end{array}
  \right.
\end{equation}
of the diverging resistance terms at the actual gap $h$ and at a
cut-off distance $\Delta_{\gamma\delta}$. Following Nguyen and
Ladd~\cite{nguyen02}, only three independent cut-offs are used,
$\Delta_{33}\equiv\Delta_\mathrm{n}$ for the coupling of normal
translation and force,
$\Delta_{44}=\Delta_{45}=\Delta_{54}=\Delta_{55}\equiv\Delta_\mathrm{r}$
for the coupling of angular velocities and torques, and
$\Delta_\mathrm{t}$ for all further resistances. Suitable values have
to be found empirically depending on $\tau$~\cite{nguyen02}.

For spheroids, the quadratic coefficients $1/(2S_1(p))$ and
$1/(2S_2(p))$ are known already from \eqnref{lub:curvature-radii}
where $p$ is defined by \eqnref{lub:spheroid} and the known surface
position $\mathbf{y}^*$. The symmetry of the particles causes
$\Gamma_1=\Gamma_3=0$. The remaining cubic coefficients
\begin{eqnarray}
  \Gamma_0(p) & = &
  \frac
  {(R_\parallel^2-R_\perp^2)R_\parallel R_\perp\sin p\cos p}
  {2s_1^6(p)}\\
  \Gamma_2(p) & = &
  \frac
  {(R_\parallel^2-R_\perp^2)R_\parallel\sin p\cos p}
  {2R_\perp s_1^4(p)}
\end{eqnarray}
are the result of an expansion of \eqnref{lub:spheroid} up to third
order in $x_1$ and $x_2$.

The method can be applied to more complex particle shapes provided
that the minimum gap, the principal curvatures, and the cubic
coefficients can be determined. The extension to mixtures of particles
of different shape or dimension can complicate the implementation to
some extent but is no further problem if an algorithm to find the
minimum separation exists, as the other parameters are properties of a
single particle. The most serious limitations are the requirement of
smooth surfaces that allow for a third-order expansion in every point
and that the surfaces must be such that for $h=0$ they would touch in
only one point~\cite{cox74}. This allows, for example, the application
to one sufficiently smoothly capped cylinder interacting with a sphere
but not to two such cylinders in parallel orientation side by
side. The occurrence of $\sqrt{\lambda_1\lambda_2}$ in the denominator
of all matrix elements \eqnref{app:first-element} to
\eqnref{app:last-element} leads to divergence if at least one of the
curvature eigenvalues defined in the appendix is zero and resembles
the second limitation directly. Taking the finite particle length for
the maximum radius of curvature, as suggested by Butler and
Shaqfeh~\cite{butler02}, might be a viable solution to this
problem. Unfortunately, the curvature of the popular spherocylinder is
still discontinuous at the transition line between the cylinder and
the hemispherical caps which might cause numerical difficulties.

Knowing the directions along which $x_1$, $x_2$, and $x_3$ are
measured, the transformation of $\bar{\mathbf{V}}_i$,
$\bar{\boldsymbol{\Omega}}_i$, $\bar{\mathbf{F}}_i$, and
$\bar{\mathbf{T}}_i$ to and from $\mathbf{V}_i$,
$\boldsymbol{\Omega}_i$, $\mathbf{F}_i$, and $\mathbf{T}_i$ in the
particle-independent coordinate system in which the integration is
carried out is straightforward to achieve. For rigid bodies the
relations with the center of mass velocities $\mathbf{v}_i$ and
$\boldsymbol{\omega}_i$ of a particle then read
\begin{eqnarray}
  \mathbf{V}_i
  & = &
  \mathbf{v}_i
  +
  \boldsymbol{\omega}_i\times(\mathbf{y}^*_i-\mathbf{r}_i)\\
  \text{and\quad}\mathbf{\Omega}_i
  & = &
  \boldsymbol{\omega}_i
\end{eqnarray}
and the resulting force and torque on the particle are
\begin{eqnarray}
  \mathbf{f}_i
  & = &
  \mathbf{F}_i\\
  \text{and\quad}\mathbf{t}_i
  & = &
  \mathbf{T}_i
  +
  (\mathbf{y}^*_i-\mathbf{r}_i)\times\mathbf{F}_i
  \text{ .}
\end{eqnarray}

\subsection{Treatment of particle contact}

In practice, the surface-normal lubrication interactions do not
suffice to prevent particle contact, especially in dense systems. The
lubrication interactions therefore are often reported to be clipped at
a specific value to avoid numerical instabilities~\cite{komnik04,
  hyvaluoma05, macmeccan09, clausen10b} and a short-range repulsive
force is added to act against unphysical clustering or even overlap of
particles~\cite{sierou04, hyvaluoma05, kromkamp06, macmeccan09,
  clausen10b, bertevas10, gunther13}. To limit the forces and torques
resulting from the lubrication corrections to finite values, the gap
entering \eqnref{lub:resistance-correction} is not allowed to be
smaller than a short-range cut-off $h_\mathrm{c}$ independently from
the actual gap $h$. For $h<h_\mathrm{c}$ a Hookean repulsive force
\begin{equation}\label{eq:lub:frep}
  F_\mathrm{r}(h)
  =
  \epsilon_\mathrm{c}
  \left\{
    \begin{array}{l@{\quad\text{for }}l}
      (h_\mathrm{c}-h) & 0<h<h_c\\
      h_\mathrm{c} & h\le0\\
    \end{array}
  \right.
\end{equation}
with a stiffness $\epsilon_\mathrm{c}$ is applied along the
$x_3$-direction which itself is limited to the magnitude achieved at
$h=0$. Effectively, this implementation of a short-range repulsion is
identical to the one applied by Kromkamp \textit{et
  al.}~\cite{kromkamp06} who refer to Ball and
Melrose~\cite{ball95}. Here, it is chosen solely because of its
simplicity. Alternative phenomenological approaches can be found in
the literature~\cite{sierou02,hyvaluoma05,clausen10b,metzger13}.
Repulsion forces founded in elastic theory~\cite{frijters12,gunther13}
or even more elaborate contact descriptions~\cite{thornton13} can be
employed if the objective is to capture non-hydrodynamic interactions
of particles. The complete lubrication correction and contact model
reads
\begin{eqnarray}\label{eq:lub:model}
  \lefteqn{\left(
      \begin{array}{c}
        \bar{\mathbf{F}}_i\\
        \bar{\mathbf{T}}_i
      \end{array}
    \right)}\nonumber\\
  & = & \left\{
    \begin{array}{l@{\quad\text{for }}l}
      \mu\tilde{\mathbf{K}}(h)
      \left(
        \begin{array}{c}
          \bar{\mathbf{V}}_j-\bar{\mathbf{V}}_i\\
          \bar{\boldsymbol{\Omega}}_j-\bar{\boldsymbol{\Omega}}_i
        \end{array}
      \right)
      &
      h>h_\mathrm{c}\\
      \mu\tilde{\mathbf{K}}(h_\mathrm{c})
      \left(
        \begin{array}{c}
          \bar{\mathbf{V}}_j-\bar{\mathbf{V}}_i\\
          \bar{\boldsymbol{\Omega}}_j-\bar{\boldsymbol{\Omega}}_i
        \end{array}
      \right)
      -
      F_\mathrm{r}(h)\hat{\mathbf{E}}_3
      &
      h\le h_\mathrm{c}\\
    \end{array}
  \right.
  \nonumber\\
  & = &
  -
  \left(
    \begin{array}{c}
      \bar{\mathbf{F}}_j\\
      \bar{\mathbf{T}}_j
    \end{array}
  \right)\text{ ,}
\end{eqnarray}
with $\hat{\mathbf{E}}_3$ being the unit $6$-vector connected with the
$x_3$-direction of the force $\bar{\mathbf{F}}_i$.

The above treatment introduces two additional parameters
$h_\mathrm{c}$ and $\epsilon_\mathrm{c}$. When modeling a physical
suspension, it can be interpreted as modeling ``residual Brownian
forces or particle roughness''~\cite{sierou04} and the parameters can
be used to control the strength of non-hydrodynamic effects. When
targeting the theoretical model system of smooth particles in the
absence of Brownian motion, care must be taken to ensure that the
influence of the non-hydrodynamic contact modeling is sufficiently
small to be neglected. Then the results of a simulation do not depend
on the exact values of the parameters $h_\mathrm{c}$ and
$\epsilon_\mathrm{c}$. Since in this view, $h_\mathrm{c}$ and
$\epsilon_\mathrm{c}$ are purely numerical parameters they should not
be rescaled as physical quantities when a simulation is transformed
from one spatial resolution to another. Instead it appears reasonable
to keep $h_\mathrm{c}$ constant when measured in lattice units so the
lubrication corrections are applied down to smaller relative gaps
$h/R$ at higher resolution $R$. From \eqnref{lub:frep} it is easily
seen that the maximum repulsive force is
$\epsilon_\mathrm{c}h_\mathrm{c}$. In order to prevent particle
overlap it should scale as the viscous forces, estimated as $6\pi\mu
R^2\dot{\gamma}$ for flows with a shear rate
$\dot{\gamma}$~\cite{sierou04,bertevas10}. Since in this work $\tau=1$
is assumed, a transformation from one resolution to another does not
change the viscosities and also $R^2\dot{\gamma}$ must stay constant
as it is proportional to the particle Reynolds number
$\mathrm{Re}_\mathrm{p}=4R^2\dot{\gamma}/\nu$. Thus
$\epsilon_\mathrm{c}$ should be kept constant then as well.

Also the present implementation of a link-based lubrication model as
in \eqnref{methods:dlub} employs a, then link-wise, clipping of the
lubrication force equivalent to the one in \eqnref{lub:model} and a
short-range repulsive force with a functional dependency on the gap
width identical to \eqnref{lub:frep}. Since in this case the gap is
measured link-wise and the force is not applied once per particle
along the surface-normal direction but once per link along the link
direction, separate values for the now link-wise parameters
$\epsilon_\mathrm{c}^*$ and $h_\mathrm{c}^*$ are required in
general. The link-wise correction forces further correspond to
stresses, therefore $\epsilon_\mathrm{c}^*h_\mathrm{c}^*$ should scale
as $6\pi\mu\dot{\gamma}$~\cite{clausen10b} to prevent overlap and
$\epsilon_\mathrm{c}^*$ requires appropriate rescaling upon changing
the resolution if $h_\mathrm{c}^*$ is kept constant.

\section{Validation and comparison to other models}\label{ch:results}

\subsection{Pair interactions of spheres}

For spheres, the lubrication corrections developed here are equivalent
to the ones of Nguyen and Ladd~\cite{nguyen02} who, however, present
detailed validation only for the interaction of a single sphere and a
planar wall. The accuracy of the present method is studied for the
more frequent interaction of two identical spheres. This setup is
particularly suited for validation since good analytical
approximations exist~\cite{jeffrey84}. For their link-wise model, Ding
and Aidun~\cite{ding03} show only the normal force between spheres
plotted on a linear scale which makes it hard to appreciate the actual
accuracy of the method, while the analysis of
Clausen~\cite{clausen10b} is performed for a resolution of the
particle radius with $R=10$ lattice sites that for non-deformable
particles seems unnecessarily large. Therefore, also a comparison with
the link-wise model at different $R$ follows.

In all tests one particle is placed with a random offset with respect
to the lattice near the center of a closed cubic box which imposes a
no-slip condition on the fluid velocity. In a preliminary study, the
size of the box is chosen to be $15R$ as a viable compromise between
computational costs and the undesired enhancement of the
single-particle resistances by interaction with the wall which decays
only slowly with increasing box size. Another particle is placed at a
random position with a prescribed gap $h$ between both surfaces. For
simplicity, the direction from the first to the second particle is
referred to as $w_3$ and two further directions $w_1$ and $w_2$ are
chosen randomly but mutually orthogonal to form a right-handed
system. While the first particle is held fixed, the second particle is
forced to either translate with a constant velocity $v$ along $w_1$ or
$w_3$ or to rotate with an angular velocity $\omega$ about $w_1$. The
magnitude of the velocities is chosen to result in a fixed Reynolds
number $\mathrm{Re}=vR/\nu=\omega R^2/\nu=6\times10^{-8}$ with
$\nu=1/6$ for all $R$. As was tested by increasing $\mathrm{Re}$ by a
factor of $10$, no inertial effects are visible. The simulations are
allowed to run for several $10^3$ LB steps until a steady state is
reached. The motion of the particle during this time is negligible and
no changes in the discretization are observed in general. Apart from
effects of discretization and confinement which are not expected to
diverge for decreasing $h$, symmetry arguments dictate that a
translation along $w_1$ and $w_3$ results in a force in the respective
opposite direction, a rotation about $w_1$ results in a torque in the
opposite direction, and additionally, translation and rotation along
$w_1$ results in a torque about $w_2$ and a force along $-w_2$,
respectively. For each $R$, the magnitude of these forces and torques
is averaged over at least $15$ random configurations with the same
value of $h$. Since the particle motion is prescribed, there is no
need for a short-range lubrication cut-off and repulsion, thus
$h_\mathrm{c}=\epsilon_\mathrm{c}=0$ is set. From comparison with the
results by Jeffrey and Onishi~\cite{jeffrey84} suitable long-range
cut-offs at $\tau=1$ for all $R$ are found to be
\begin{equation}\label{eq:results:cutoffs}
  \Delta_\mathrm{n}=\frac{2}{3}
  \text{ ,\quad}
  \Delta_\mathrm{t}=\frac{1}{2}
  \text{ ,\quad and\quad}
  \Delta_\mathrm{r}=\frac{1}{4}
  \text{ .}
\end{equation}
The relation to the cut-offs $h_N$, $h_T$, and $h_R$ suggested by
Nguyen and Ladd~\cite{nguyen02} is not surprising:
$\Delta_\mathrm{n}=h_N$ and $\Delta_\mathrm{t}=h_T$ but
$\Delta_\mathrm{r}\not=h_R$ because the cut-offs in
Ref.~\cite{nguyen02} relate to the resistance functions of the spheres
themselves while the cut-offs here relate to the resistances of the
surfaces at the gap. Thus, in the lubrication correction of the torque
experienced due to rotation about $w_1$, both $\Delta_\mathrm{t}$ and
$\Delta_\mathrm{r}$ are involved while in Ref.~\cite{nguyen02} it is
only $h_R$.

\begin{figure}
  \includegraphics[width=1.0\columnwidth]
  {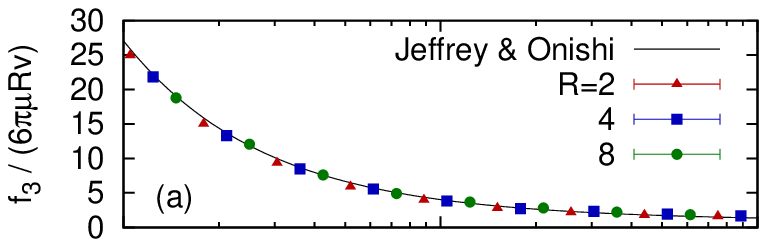}\\[-7ex]
  \includegraphics[width=1.0\columnwidth]
  {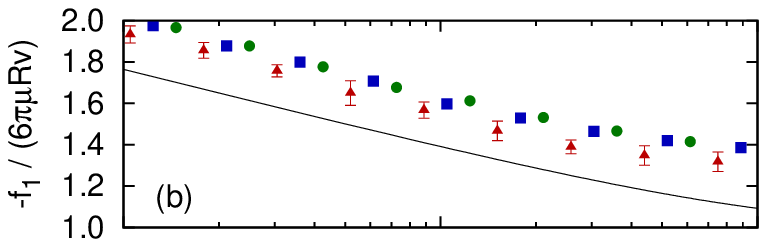}\\[-7ex]
  \includegraphics[width=1.0\columnwidth]
  {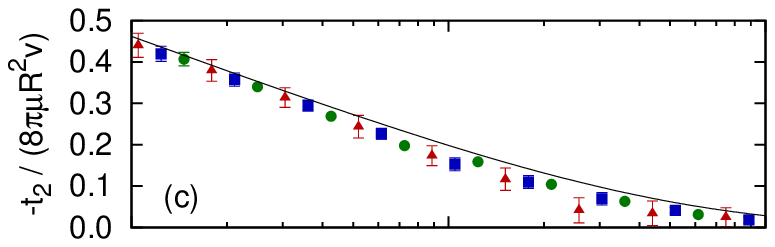}\\[-7ex]
  \includegraphics[width=1.0\columnwidth]
  {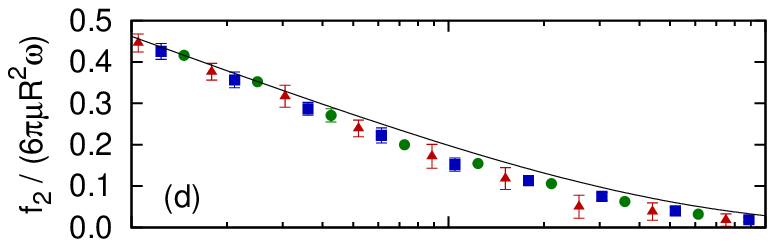}\\[-7ex]
  \includegraphics[width=1.0\columnwidth]
  {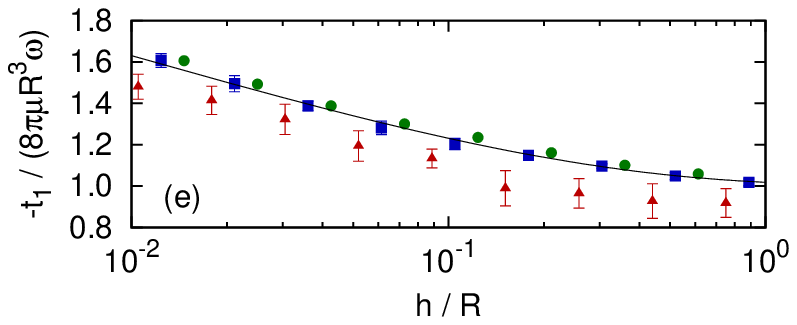}
  \caption{\label{fig:results:spheres-club}(Color online)
    Non-dimensional resistances for one of two identical spheres of
    radius $R$ aligned along direction $w_3$ at a normalized gap
    $h/R$. Compared are simulations with the contact-based
    lubrication model at different resolutions of $R$ in lattice units
    (symbols) with the series developed by Jeffrey and
    Onishi~\cite{jeffrey84} (lines). (a) shows the force along $w_3$
    due to a translation with velocity $v$ in the opposite direction,
    (b) the force in the perpendicular direction $-x_1$ due to a
    translation along $w_1$, (c) the torque about $w_2$ due to the
    same translation, (d) the force along $-x_2$ due to a rotation
    about $w_1$ with angular velocity $\omega$, and (e) the resulting
    torque about $-x_1$. Error bars quantify the standard deviation
    obtained from at least $15$ random configurations and are drawn
    only where larger than the symbol itself.}
\end{figure}

\picref{results:spheres-club} displays the resistance functions
related to the particle motions explained above at resolutions $R=2$,
$4$, and $8$ resulting from the choice of lubrication cut-offs given
in \eqnref{results:cutoffs}. The striking similarity of
\picref{results:spheres-club}(c) and (d) can be explained by the
Lorentz reciprocal theorem~\cite{kim05}. The data in
\picref{results:spheres-club}(b) and (c), and the data in (d) and (e)
are taken from the same simulations. Generally, an offset with respect
to the solution by Jeffery and Onishi~\cite{jeffrey84} is observed
which in each figure is roughly constant for a given $R$. These
offsets can be attributed to two effects: the analytical solution
considers two particles in an infinite volume of fluid while the
simulations are performed within a finite box. The interactions with
the walls lead to an enhancement of the single-particle resistances,
best seen in \picref{results:spheres-club}(b) for large $h/R$. Second,
the effective hydrodynamic radii of particles modeled by the momentum
exchange method are known to differ from the input radii $R$,
especially at small $R$, which explains the more or less severe
underestimation of all resistances for $R=2$. The comparably large
statistical errors in the data for the smallest radius must be
attributed to discretization effects. All these observations could be
made for single particles as well and it is not within the scope of
lubrication corrections to alleviate the shortcomings but only to
produce a smooth increase of the resistances with decreasing
separation parallel to the theoretical solution. The lubrication model
achieves this objective to very good accuracy at $R=4$ and $8$ and,
taking into account the larger discretization errors, even at $R=2$.

\begin{figure}
  \includegraphics[width=1.0\columnwidth]
  {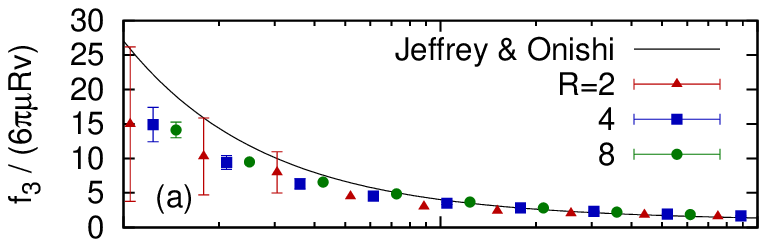}\\[-7ex]
  \includegraphics[width=1.0\columnwidth]
  {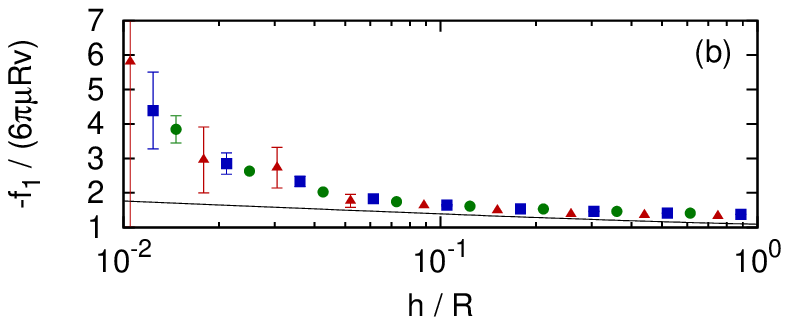}
  \caption{\label{fig:results:spheres-dlub}(Color online) Normalized
    resistances as in \picref{results:spheres-club} obtained from
    simulations with link-based lubrication corrections. For brevity,
    only the force due to a (a) normal or (b) tangential translation
    is shown, the other singular resistances appear to be
    qualitatively similar to case (b).}
\end{figure}

\begin{figure}
  \includegraphics[width=1.0\columnwidth]
  {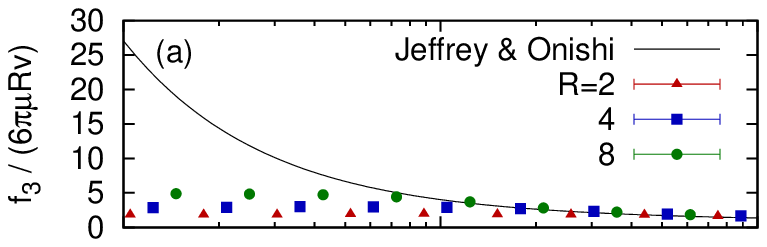}\\[-7ex]
  \includegraphics[width=1.0\columnwidth]
  {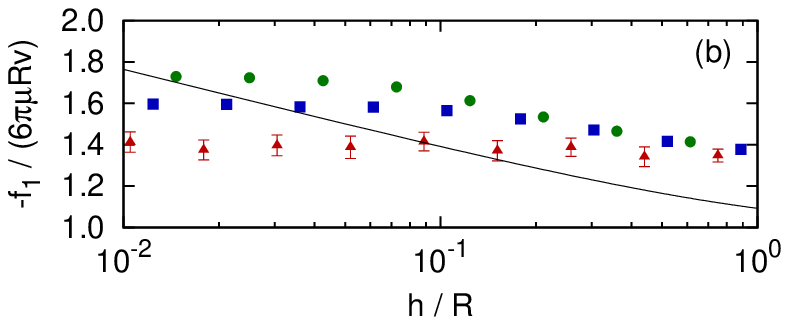}
  \caption{\label{fig:results:spheres-nolub}(Color online) Normalized
    resistances as in \picref{results:spheres-dlub} obtained from
    simulations without lubrication corrections.}
\end{figure}

As expected from the work of Clausen~\cite{clausen10b}, the link-wise
model leads to an under-prediction of the normal force $f_3$ as
displayed in \picref{results:spheres-dlub}(a) while the divergence of
the tangential force $f_1$ in \picref{results:spheres-dlub}(b) is
clearly over-predicted. The other non-normal singular forces and
torques $t_2$, $f_2$, and $t_1$ are over-estimated in a qualitatively
similar way at small separations. A special treatment of short-range
contacts is disabled in these simulations by setting
$h_\mathrm{c}^*=\epsilon_\mathrm{c}^*=0$. Since the link-based model
relies on discrete lattice links to compute lubrication corrections,
severe noise is caused by an amplification of discretization errors at
small distances. This effect is strongest for $R=2$ but visible also
by the error bars obtained for the higher resolutions in
\picref{results:spheres-dlub}. Finally, \picref{results:spheres-nolub}
shows the resistances to normal and tangential translation without
lubrication corrections. Again, the other non-normal resistances
appear to be qualitatively similar to
\picref{results:spheres-nolub}(b). It is clear that, without
lubrication modeling, the singular behavior is captured only partially
depending on the lattice resolution. In the case of $R=2$, an increase
is hardly visible for gaps $h/R<1$. In Clausen's modified link-based
model~\cite{clausen10b} the under-prediction of $f_3$ appears to be
reduced while all non-normal resistances remain uncorrected. At best,
this model can perform as the theoretical solution for $f_3$, which
is, as can be seen in \picref{results:spheres-club}(a), closely
approximated by the contact-based method, and comparable to
\picref{results:spheres-nolub}(b) for all other terms. The importance
of non-normal interactions in dynamic simulations of many particles
remains to be examined below.

\subsection{Pair interactions of spheroids}

\begin{figure*}
  \includegraphics[width=1.0\columnwidth]
  {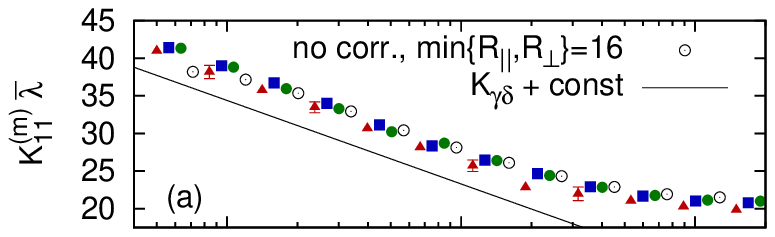}\hfill%
  \includegraphics[width=1.0\columnwidth]
  {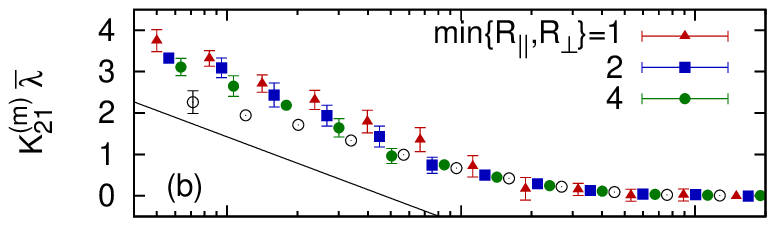}\\[-7ex]
  \includegraphics[width=1.0\columnwidth]
  {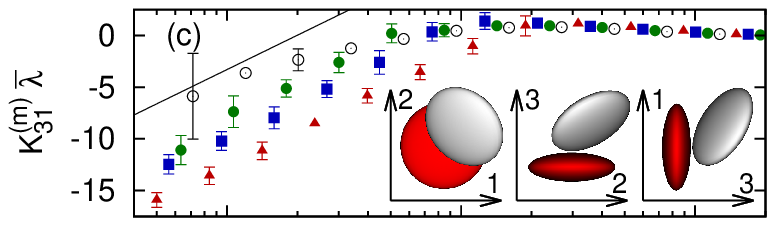}\hfill%
  \includegraphics[width=1.0\columnwidth]
  {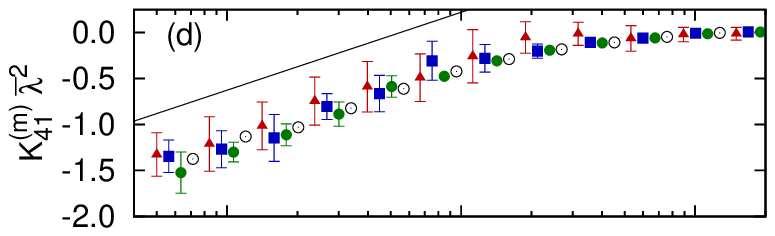}\\[-7ex]
  \includegraphics[width=1.0\columnwidth]
  {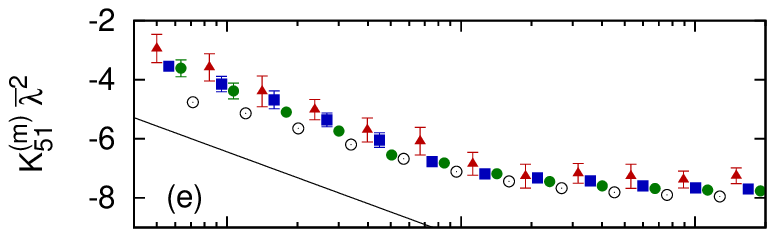}\hfill%
  \includegraphics[width=1.0\columnwidth]
  {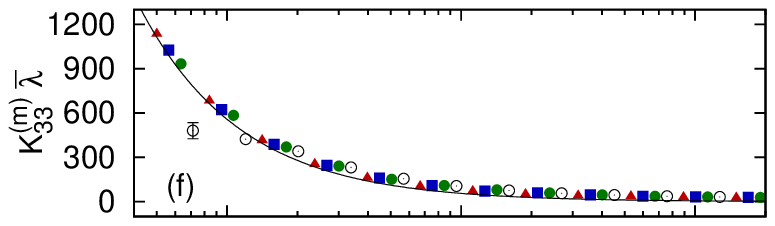}\\[-7ex]
  \includegraphics[width=1.0\columnwidth]
  {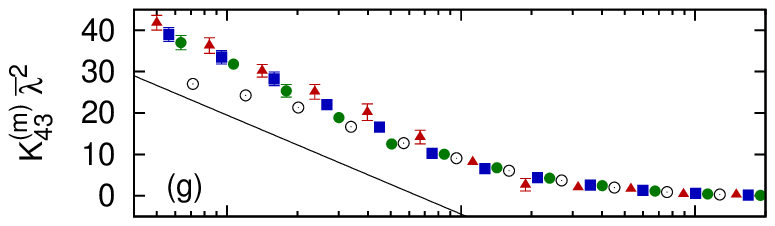}\hfill%
  \includegraphics[width=1.0\columnwidth]
  {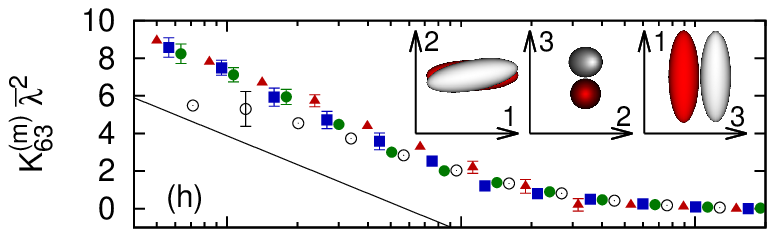}\\[-7ex]
  \includegraphics[width=1.0\columnwidth]
  {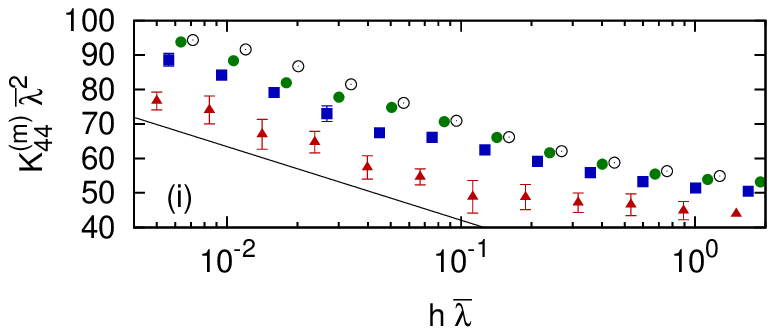}\hfill%
  \includegraphics[width=1.0\columnwidth]
  {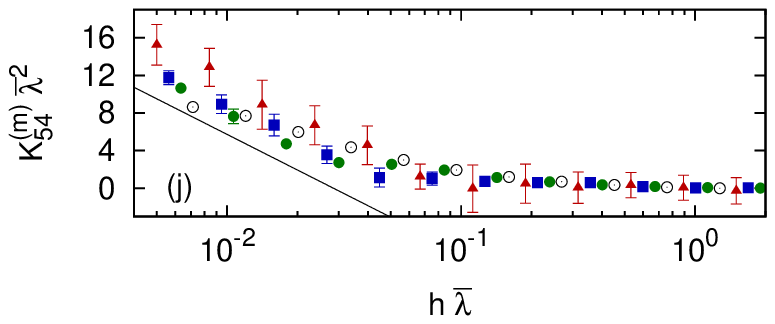}
  \caption{\label{fig:results:spheroid-resolution}(Color online) The
    measured resistances $K_{\gamma\delta}^\mathrm{(m)}$ for the ten
    independent and physically different singular types of lubrication
    interactions as function of the surface separation $h$: (a)
    tangential force, (b) perpendicular tangential force, (c) normal
    force, (d) tangential torque, and (e) perpendicular tangential
    torque due to tangential translation; (f) normal force, (g)
    tangential torque, and (h) normal torque due to normal
    translation; and (i) tangential torque and (j) perpendicular
    tangential torque due to tangential rotation. All data is obtained
    from the same configuration of oblate spheroids with aspect ratio
    $\Lambda=1/3$ except for (h) where prolates with $\Lambda=3$ are
    used. The two configurations are visualized in the insets of
    figures (c) and (h) as seen from the respective local $x_3$-,
    $x_1$-, and $x_2$-direction of the first of the two particles
    (drawn red, with center at lower coordinate position). Symbols
    refer to different resolutions of the smaller half-axis, lines to
    the respective correction term, shifted by an arbitrary
    constant. Resolutions $1$ to $4$ feature lubrication corrections,
    resolution $16$ does not for comparison. The combined mean
    curvature $\bar{\lambda}$ serves for non-dimensionalization.}
\end{figure*}

To validate the lubrication corrections in the case of spheroidal
particles, an equivalent procedure as for spheres is followed. The
long-range cut-offs in \eqnref{results:cutoffs} found suitable for
spheres are kept. Different from the case of spheres, the validation
is performed not focusing on the particles as a whole but on the
surfaces around the points of closest approach $\mathbf{y}^*_i$ and
$\mathbf{y}^*_j$ on both spheroids. Thus, the coordinate system
$\{x_1,x_2,x_3\}$ is chosen as defined in \eqnref{lub:surface-i}
according to the local directions of principal curvature and the local
normal direction of the first particle. While the second particle is
held fixed, the first one is forced to translate and rotate in a way
that causes its surface to either translate along or rotate about one
of the coordinate axes and the resulting force and the torque with
respect to $\mathbf{y}^*_i$ is recorded in the same frame. Compared to
spheres, the parameter space is considerably increased as a
configuration is defined not only by the surface separation but also
by the particle aspect ratio and the relative orientation. The purpose
here lies mainly in demonstrating the physical correctness and the
degree of resolution independence achievable with the presented
model. For this purpose it is sufficient to exemplarily examine one
fixed particle configuration at varying distances for each diverging
element of $\mathbf{K}$. The 16 singular $K_{\gamma\delta}$ can be
reduced to the only 10 physically different cases that are displayed
in \picref{results:spheroid-resolution}: (a) tangential force $K_{11}$
(equivalent to $K_{22}$), (b) perpendicular tangential force $K_{21}$,
(c) normal force $K_{31}$ (equivalent to $K_{32}$), (d) tangential
torque $K_{41}$ (equivalent to $K_{52}$), and (e) perpendicular
tangential torque $K_{51}$ (equivalent to $K_{42}$) due to tangential
translation; (f) normal force $K_{33}$, (g) tangential torque $K_{43}$
(equivalent to $K_{53}$), and (h) normal torque $K_{63}$ due to normal
translation; and (i) tangential torque $K_{44}$ (equivalent to
$K_{55}$) and (j) perpendicular tangential torque $K_{54}$ due to
tangential rotation. \picref{results:spheroid-resolution}(a) to (e)
are generated from the same set of simulations. The same is done for
plots (f) and (g) and for (i) and (j). For (h) $K_{63}$ a
configuration of two prolate spheroids with aspect ratio $\Lambda=3$
is chosen where the axes of both particles are both perpendicular to
the line between their centers but twisted against each other by an
angle of $10^\circ$. The remaining resistances are examined for a
randomly generated configuration of two oblates $\Lambda=1/3$ that
results in a clear divergence of all elements but $K_{63}$ and that is
defined by the angles
$\angle(\mathbf{y}_i^*-\mathbf{r}_i,\hat{\mathbf{o}}_i)=177.06^\circ$,
$\angle(\mathbf{y}_j^*-\mathbf{r}_j,\hat{\mathbf{o}}_j)=82.591^\circ$,
and $\angle(\hat{\mathbf{o}}_i,\hat{\mathbf{o}}_j)=40.277^\circ$ with
the respective axes of rotational symmetry $\hat{\mathbf{o}}_i$ and
$\hat{\mathbf{o}}_j$ of the particles. The configurations are
visualized as insets of \picref{results:spheroid-resolution}(c) and
(h). The surface separation $h$ but also the resistances
$K_{\gamma\delta}$ are made dimensionless using the combined mean
curvature of both surfaces~\cite{claeys89}
\begin{equation}
  \bar{\lambda}=\frac{1}{4}
  \left[
    \frac{1}{S_1}+
    \frac{1}{S_2}+
    \frac{1}{S_1'}+
    \frac{1}{S_2'}
  \right]
  \text{ .}
\end{equation}
Again the magnitude of the velocity is chosen such that $\mathrm{Re}$
remains constant when changing the resolution. In case of the lowest
resolution $\min\{R_\parallel,R_\perp\}=1$ it is $v=10^{-8}$. The size
of the cubic simulation volume in this case is
$30\min\{R_\parallel,R_\perp\}=10\max\{R_\parallel,R_\perp\}=30$ and
is scaled according to the particle resolution. For the data at
resolutions $1$, $2$, and $4$, each symbol stands for the average of
at least $15$ independent simulations at identical relative particle
separation and orientation but a random sub-grid offset and rotation
with respect to the fluid lattice. The error bars represent the
standard deviations.

In the absence of theoretical results, simulations with lubrication
corrections at several resolutions are compared to simulations without
corrections but with a considerably increased lattice resolution
$\min\{R_\parallel,R_\perp\}=16$ where lubrication interactions can be
expected to be captured by the LB method itself already to a large
extent. For these more expensive high-resolution simulations, each
symbol in \picref{results:spheroid-resolution} corresponds to only $4$
independent samples. Additionally, the divergence behavior expected
from \eqnref{app:first-element} to \eqnref{app:last-element} is
plotted with an arbitrary offset. As a further confirmation for the
correct implementation of the method, Newton's third law in
\eqnref{lub:model} is not exploited to save computations but the
effects on both surfaces are computed independently and checked for
consistency. Indeed, the forces and torques are found to be equal but
opposite apart from small deviations attributed to the error in the
anti-parallelism of the normal directions of both surfaces according
to \eqnref{lub:convergence}.

In \picref{results:spheroid-resolution} the resistances obtained with
lubrication corrections at the larger resolutions
$\min\{R_\parallel,R_\perp\}=2$ and $4$ appear largely parallel to
each other and consistent with data from the uncorrected simulations
at resolution $16$. Of course, also at this highest resolution the
lubrication interactions as resolved by the LB method alone eventually
break down at the smallest separations and consequentially depend
strongly on the discretization of the gap. These effects are best
visible in \picref{results:spheroid-resolution}(h) and (c). Still, at
least for intermediate gaps, the resistances obtained at resolution
$16$ are well compatible with the theoretical divergence terms
$K_{\gamma\delta}$. Simulations at the lowest resolution
$\min\{R_\parallel,R_\perp\}=1$ tend to suffer more from
discretization errors and in general feature a less smooth transition
from the non-singular long-range behavior to the short-range regime
that is dominated by the $K_{\gamma\delta}$. For the chosen particle
configuration, the normal force induced by a tangential translation,
displayed in \picref{results:spheroid-resolution}(c), seems to be
particularly difficult to capture properly: for large gaps the
resistance seems identical at all resolutions but in the region where
lubrication interactions dominate an offset is visible not only at
resolution $1$ but also between resolutions $2$ and $4$, and compared
to resolution $16$ without corrections. This indicates that in this
particular case, different than assumed, the divergence of $K_{31}$ is
not resolved by the LB method down to a separation of approximately
one lattice unit. A possible explanation lies in the fact that
$K_{31}$ (given in \eqnref{app:chi13}) depends on the third order
coefficients $\Gamma_0$ and $\Gamma_2$ of both particles.
On the second particle, $\mathbf{y}^*_j$ lies near the edge (see inset
of \picref{results:spheroid-resolution}(c)) where according to
\eqnref{lub:curvature-radii} the minimum radius of curvature for an
oblate of aspect ratio $1/3$ is only $R_\parallel/3$. Even at
resolution $4$, the radius is only $4/3$ and the third-order features
of the surface cannot be expected to be resolved well at this
resolution. It is not clear why equivalent difficulties do not exist
for the normal force induced by a tangential rotation depicted in
\picref{results:spheroid-resolution}(g) that also depends on
$\Gamma_0$ and $\Gamma_2$ in the same configuration of particles. The
error bars, also of the data at resolution 16 without corrections,
suggest that $K_{31}$ in this configuration is particularly
susceptible to discretization effects. Nevertheless, at resolution
$\min\{R_\parallel,R_\perp\}=2$ the offset to higher resolution data
at small gaps does not appear exceedingly large as compared to the
results for the other resistances. In the following simulations of
spheroidal particles a length of the smaller half-axis of 2 lattice
units is regarded as minimum spatial resolution.

\subsection{Many particles in suspension: shear-induced diffusion and
  viscosity}

The initialization of densely-packed configurations of many particles
without overlap is not trivial already for spherical particles. A
growth method similar to the ones described in the
literature~\cite{clausen10,kruger12} is applied: the particles are
initially scaled down to typically only $30\,\%$ of their linear size,
so an overlap-free placement at random positions is easily
achieved. The half-axes are then slowly grown to their actual
size. During the growth, the particles are free to reorient and move
but lubrication and hydrodynamic interactions are replaced by simple
damping terms for the velocities and only the Hookean short-range
repulsion defined in \eqnref{lub:frep} is acting. Growing the
particles to dimensions slightly beyond their final size and resetting
the size afterwards assures a certain minimum separation in the
generated particle configuration.

As a first benchmark, the shear-induced self-diffusion of spherical
particles in the direction of a velocity gradient is studied. Accurate
data for supposedly purely hydrodynamically interacting spheres in
Stokes flow is reported by Sierou and
Brady~\cite{sierou04}. Lees-Edwards boundary
conditions~\cite{lorenz09b} impose a well-defined shear rate in an
otherwise periodic system with cubic dimensions to mimic an infinite
volume of fluid. Depending on the volume fraction $\Phi$, between
$800$ to $4000$ particles are modeled, which, according to
Ref.~\cite{sierou04} should allow for sufficiently reliable
results. In the case of a resolution of $R=4$, the simulation
comprises $128^3$ lattice sites. Initially, all systems are
equilibrated with the shear for a time interval of roughly
$130\dot{\gamma}^{-1}$. For each particle $i$, the trajectory $x_i(t)$
in velocity gradient direction is recorded during another
$630\dot{\gamma}^{-1}$ and the mean-square displacement $\langle\Delta
x^2(\Delta t)\rangle=\langle(x_i(t+\Delta t)-x_i(t))^2\rangle_{i,t}$
during a time interval $\Delta t$ is computed as an average over all
particles $i$ and starting times $t$.

\begin{figure}
  \includegraphics[width=\columnwidth]
  {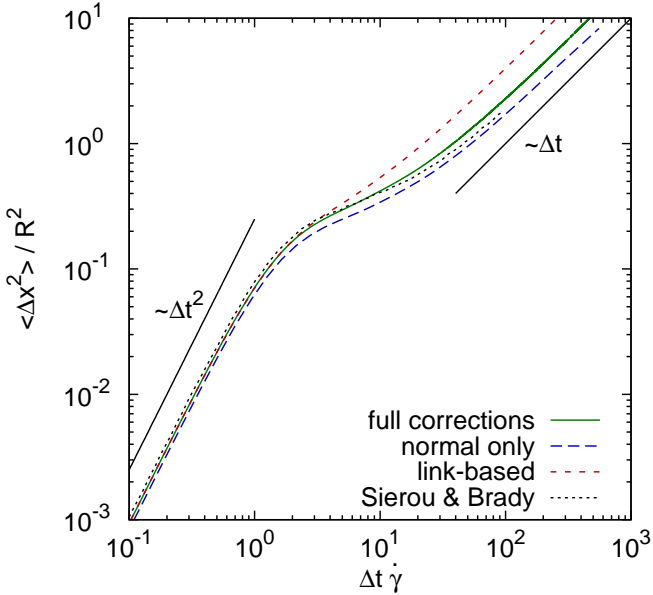}
  \caption{\label{fig:results:diffusion-msd}(Color online)
    Shear-induced self-diffusion in suspensions of spheres with radius
    $R=4$ as obtained with the full contact-based corrections, the
    same with non-normal corrections disabled, and the link-based
    approach. The mean square displacement in velocity gradient
    direction at a solid volume fraction of $\Phi=0.2$ is
    shown. Results from accelerated Stokesian dynamics simulations by
    Sierou and Brady~\cite{sierou04} are plotted for comparison.}
\end{figure}

\begin{figure}
  \includegraphics[width=\columnwidth]
  {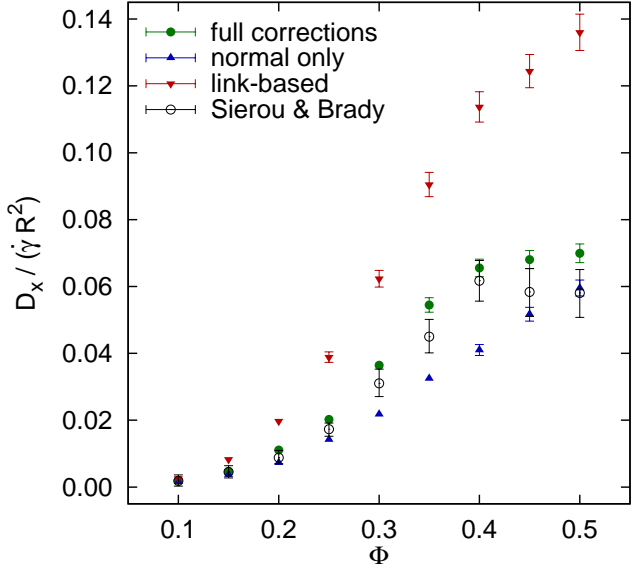}
  \caption{\label{fig:results:diffusion-d}(Color online) Shear-induced
    self-diffusion as in \picref{results:diffusion-msd}. The diffusion
    coefficients in velocity gradient direction computed from the
    mean-square displacement as shown in
    \picref{results:diffusion-msd} are displayed as a function of the
    volume fraction $\Phi$. Results from accelerated Stokesian
    dynamics simulations by Sierou and Brady~\cite{sierou04} are shown
    for comparison. The error bars represent an estimated statistical
    error of $4\,\%$ and are drawn only where larger than the
    corresponding symbol.}
\end{figure}

\picref{results:diffusion-msd} compares the mean-square particle
displacement $\langle\Delta x^2\rangle$ for $\Phi=0.2$ as a function
of the time interval $\Delta t$ as obtained from the contact-based
method presented above, the link-wise method briefly sketched in
section~\chref{methods:link-based}, and the contact-based method with
all non-normal lubrication corrections disabled with the results by
Sierou and Brady~\cite{sierou04}. As expected, over short times
$\Delta t$ the particle velocities are self-correlated and a scaling
$\langle\Delta x^2\rangle\sim\Delta t^2$ is found. For large $\Delta
t$ hydrodynamic interactions lead to decorrelation and the relation
becomes diffusive with $\langle\Delta x^2\rangle\sim\Delta t$. All
lubrication corrections capture the short-time behavior consistently
and show diffusion at long times. While, however, in the two
contact-based simulations, the magnitude of the diffusion is only
moderately under- and over-predicted as compared to the literature,
the link-wise corrections lead to a considerable over-prediction by
roughly $100\,\%$. This observation is made for the full range of
volume fractions when comparing the resulting diffusion coefficients
$D_x$, obtained for large $\Delta t$ from fits $\langle\Delta
x^2\rangle=2D_x\Delta t+\mathrm{const}$, as it is done in
\picref{results:diffusion-d}. By repeating the simulations using four
different random starting configurations for $\Phi=0.2$ and $0.4$, the
statistical errors are estimated to be $\Delta D_x=0.0004$ and
$0.002$, respectively, or $4\,\%$ and $2\,\%$ of the respective $D_x$
itself. In \picref{results:diffusion-d}, a relative statistical error
of $4\,\%$ is assumed for all data. In the case where the non-normal
interactions are disabled, $D_x$ continues to grow as a function of
$\Phi$ at $\Phi>40\,\%$ where all other datasets show a plateau or at
least a significantly reduced slope. In general, the full
contact-based corrections are most consistent with literature
data. The remaining discrepancy might be caused by finite size effects
or by small deviations in the effective hydrodynamic particle
radii. The finite particle Reynolds number
$\mathrm{Re}_\mathrm{p}=4R^2\dot{\gamma}/\nu=0.24$ seems sufficiently
small to justify the comparison with Stokesian dynamics
simulations. In their recent publication~\cite{yeo13}, Yeo and Maxey,
using the lubrication-corrected force coupling method, find values of
$D_x$ almost identical to those in Ref.~\cite{sierou04} even at
(according to the above definition) $Re_\mathrm{p}=0.4$.

\begin{table}
  \centering
  \begin{tabular}{lrll@{\hspace*{0.8em}}ll@{\hspace*{0.8em}}l}
    \hline\hline
    $h_\mathrm{c}^{(*)}$ & $\epsilon_\mathrm{c}^{(*)}$ & model &
    $R$ & $D_x(\Phi\!=\!0.2)$ & $D_x(\Phi\!=\!0.4)$\\
    \hline
    0.01  & 100  & full & 4 & $0.011\pm0.0004$ & $0.066\pm0.002$\\
    0.1   & 100  & full & 4 & $0.012$ & $0.058$\\
    0.001 & 100  & full & 4 & $0.010$ & $0.068$\\
    0.01  & 50   & full & 4 & $0.011$ & $0.064$\\
    0.01  & 1000 & full & 4 & $0.011$ & $0.063$\\
    0.01  & 100  & normal only & 4 & $0.0073$ & $0.041$\\
    0.04  & 100  & link-based  & 4 & $0.020$  & $0.11$\\[2ex]
    0.01  & 100  & full & 2 & $0.011$ & $0.062$\\
    0.01  & 100  & normal only & 2 & $0.0070$ & $0.035$\\
    0.04  & 400  & link-based  & 2 & $0.021$  & $0.10$\\[2ex]
    0.01  & 100  & full & 8 & $0.011$ & $0.068$\\
    0.01  & 100  & normal only & 8 & $0.0085$ & $0.050$\\
    0.04  & 25  & link-based  & 8 & $0.019$ & $0.14$\\
    \hline\hline
  \end{tabular}
  \caption{\label{tab:results:diffusion-spheres}Shear-induced self-diffusion
    $D_x$ in the velocity-gradient direction for spherical particles at
    volume fractions $\Phi=0.2$ and $0.4$ using different lubrication models,
    resolutions $R$, and short-range numerical parameters $h_\mathrm{c}$ and
    $\epsilon_\mathrm{c}$ ($h_\mathrm{c}^*$ and $\epsilon_\mathrm{c}^*$ in case of
    link-based lubrication corrections). The particle Reynolds number is
    $\mathrm{Re}_\mathrm{p}=4R^2\dot{\gamma}/\nu=0.24$. Statistical error
    estimates are computed from simulations with different random seeds for
    $h_\mathrm{c}=0.01$, $\epsilon_\mathrm{c}=100$, and $R=4$ exemplarily.}
\end{table}

Obviously, the influence of the free short-range parameters
$h_\mathrm{c}^{(*)}$ and $\epsilon_\mathrm{c}^{(*)}$ on shear-induced
diffusion needs to be examined. In the present simulations,
$h_\mathrm{c}=0.01$ and $\epsilon_\mathrm{c}=100$ is chosen for the
two contact-based cases and $h_\mathrm{c}^*=0.04$ and
$\epsilon_\mathrm{c}^*=100$ for the link-wise corrections. Measured on
the scale of viscous forces or stresses, this corresponds to maximum
repulsions of $h_\mathrm{c}\epsilon_\mathrm{c}/(6\pi\mu
R^2\dot{\gamma})\approx30$ and
$h_\mathrm{c}^*\epsilon_\mathrm{c}^*/(6\pi\mu\dot{\gamma})\approx2000$,
respectively. In \tabref{results:diffusion-spheres}, diffusion
coefficients for volume fractions $\Phi=0.2$ and $0.4$ are shown at
varying $h_\mathrm{c}$ and $\epsilon_\mathrm{c}$ for different
lubrication models at resolution $R=4$. Additionally, data for
resolutions $R=2$ and $8$ is shown where $h_\mathrm{c}$ and
$h_\mathrm{c}^*$ are kept fixed and $\epsilon_\mathrm{c}$ and
$\epsilon_\mathrm{c}^*$ are chosen such that the respective
dimensionless maximum repulsion is the same as for $R=4$. The numbers
do not seem very sensitive to even large changes in the short-range
parameters of $h_\mathrm{c}$ to $50$ or $1000$ and of
$\epsilon_\mathrm{c}$ to $0.1$ or $0.001$, the largest relative change
being a reduction of about $12\,\%$ at $\Phi=0.4$ for
$h_\mathrm{c}=0.1$. Interestingly, the same change effects an increase
of $D_x$ at $\Phi=0.2$. Values $\epsilon_\mathrm{c}<50$ do not suffice
to prevent particles from overlapping in the present simulations. It
is surprising that for $\Phi=0.2$, diffusion remains unchanged when
reducing the resolution to only $R=2$ or doubling it to $R=8$ and also
at $\Phi=0.4$ there hardly is a significant dependency on $R$. It
appears that shear-induced self-diffusion at the volume fractions
considered is determined rather by short-range lubrication
interactions than by hydrodynamic interactions acting over larger
length scales that could be resolved by pure LB simulations at
practical resolutions of the particle radii. The results for purely
normal and link-wise lubrication corrections remain qualitatively
unchanged when varying the resolution: without non-normal corrections,
diffusion is reduced by $23$ to $44\,\%$, the link-wise model leads to
an increase between $61$ and $106\,\%$, depending on resolution and
volume fraction. Unlike the full model, the two other methods show a
significant dependency on the resolution except for the link-based
model at $\Phi=0.2$.

There seems to be less freedom in the choice of the short range
parameters $h_\mathrm{c}^*$ and $\epsilon_\mathrm{c}^*$ associated
with the link-based lubrication model than for the contact-based
model. Significant deviations from the combination
$h_\mathrm{c}^*=0.04$ and $\epsilon_\mathrm{c}^*=100$ at $R=4$ are
often found to lead to numerical instabilities even when the time step
for the particle update is reduced. Still, the over-estimation of
diffusion by the link-wise model is too strong to be attributed solely
to the treatment of particle contacts. The stability problems might be
caused by the relatively large discretization errors of the method
that are visible in \picref{results:spheres-dlub}. A related issue
consists in contacts with large curvatures of the involved particle
surfaces staying undetected due to an insufficient resolution of the
surfaces by lattice links. In consequence, unphysically close approach
or even overlap of particles is possible. If the discretization
changes because the involved particles are moving with respect to the
lattice, the contact can get resolved and large repulsive forces and
changes in the particles' resistances can emerge suddenly. Though such
events are rare and thus are not expected to affect the observables in
a large system they can cause a simulation to crash. At large particle
resolutions, as chosen by Clausen~\cite{clausen10b}, this problem is
less likely to occur.

As a second benchmark, the shear viscosity of suspensions is
considered. For viscosity computation, the method successfully applied
before~\cite{janoschek10} is used: different from the full
Lees-Edwards boundary conditions employed above, the particles are now
prevented from crossing the sheared boundaries at which the shear
stress $\sigma$ is computed. To exclude boundary effects, the shear
rate $\dot{\gamma}$ and the actual particle volume fraction $\Phi$ are
computed only from the particles in the central half of the
system. Depending on the volume fraction and particle aspect ratio,
the total simulations comprise between about $600$ and $9000$
particles which is more than sufficient to obtain reliable
results~\cite{sierou02}. At the beginning of a simulation, each system
is allowed a time of the order of $20\dot{\gamma}^{-1}$ for
equilibration. An equally long consecutive interval of time is used
for data accumulation. The statistical errors are estimated from the
fluctuations of $\dot{\gamma}$ and $\sigma$ over time that are
propagated into the relative suspension viscosity
$\mu_\mathrm{r}=(\sigma/\dot{\gamma})/\mu$.

\begin{figure}
  \includegraphics[width=\columnwidth]{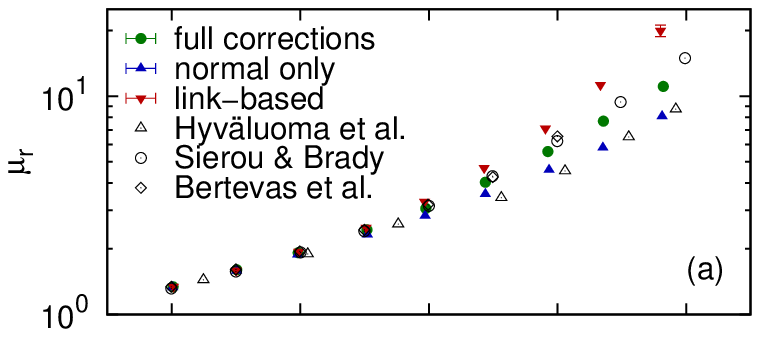}\\[-8ex]
  \includegraphics[width=\columnwidth]{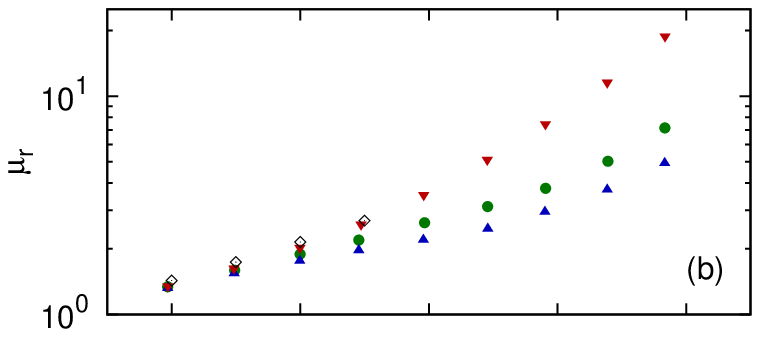}\\[-8ex]
  \includegraphics[width=\columnwidth]{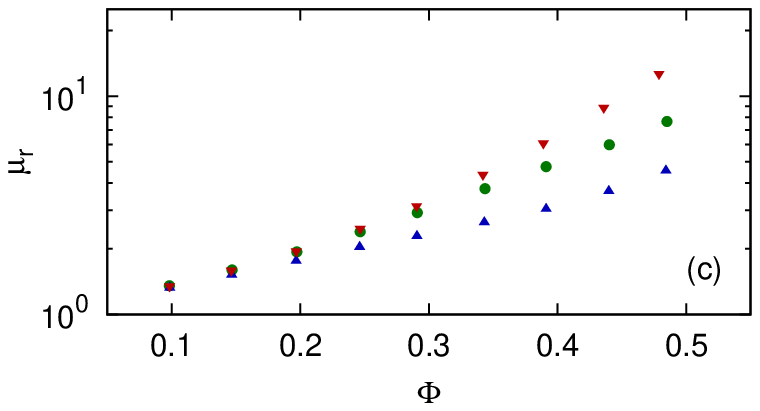}
  \caption{\label{fig:results:viscosity}(Color online) Relative
    suspension viscosity $\mu_\mathrm{r}$ in dependence on the solid
    volume fraction $\Phi$ as obtained with the full contact-based
    corrections, the same with non-normal corrections disabled, and
    the link-based approach for (a) spheres with radius $R=4$, (b)
    oblates with $R_\parallel=2$, $R_\perp=6$ (aspect ratio
    $\Lambda=1/3$), and (c) prolates with $R_\parallel=6$, $R_\perp=2$
    ($\Lambda=3$). For spheres, data is compared with results from LB
    simulations with only normal corrections at $R=6$ by Hyv\"aluoma
    \textit{et al.}~\cite{hyvaluoma05}, accelerated Stokesian dynamics
    simulations by Sierou and Brady~\cite{sierou02} and the method by
    Bertevas \textit{et al.}~\cite{bertevas10}. Also for oblates, data
    for the least aspherical aspect ratio $\Lambda=0.3$ studied by
    Bertevas \textit{et al.}~\cite{bertevas10} is shown. Error bars
    for original data are drawn only where larger than the
    corresponding symbol.}
\end{figure}

The viscosities thus obtained are plotted as a function of $\Phi$ in
\picref{results:viscosity}(a) in the case of spheres. The resolution
is $R=4$ and again the three different lubrication models are
compared: the contact-based model developed above, the link-based
model briefly sketched in section~\chref{methods:link-based}, and the
contact-based model with all non-normal corrections disabled. Of these
three cases, the full contact-based model clearly shows the best
consistency with the accelerated Stokesian dynamics simulations by
Sierou and Brady~\cite{sierou02} and the simpler method by Bertevas
\textit{et al.}~\cite{bertevas10} that both similarly aim at the
simulation of purely hydrodynamically interacting particles at low
Reynolds number shear flow. While the link-wise model leads to an
over-prediction of viscosities, the omission of non-normal corrections
results in erroneously low viscosities. In view of
\picref{results:spheres-dlub}(b) and \picref{results:spheres-nolub}(b)
it seems plausible to attribute these errors directly to the over- or
under-prediction of non-normal lubrication interactions. The
similarity of the data for purely normal lubrication corrections as
compared with the results by Hyv\"aluoma \textit{et
  al.}~\cite{hyvaluoma05} obtained at a somewhat higher resolution of
$R=6$ using normal lubrication corrections of the type of
\eqnref{methods:llub} confirms the validity of the LB implementation
and of the procedure for viscosity measurement employed here. It is
interesting that for $\Phi\lesssim0.2$ all three lubrication models
produce viscosities that are hard to distinguish, even on the
logarithmic scale of $\mu_\mathrm{r}$ in the figure. Apparently,
lubrication interactions, at least the non-normal ones, contribute
only weakly to the viscosity at these volume fractions.

\begin{table}
  \centering
  \begin{tabular}{lrll@{\hspace*{0.8em}}ll@{\hspace*{0.8em}}r@{$\,\pm\,$}l}
    \hline\hline
    $h_\mathrm{c}^{(*)}$ & $\epsilon_\mathrm{c}^{(*)}$ & model &
    $R$ & $\Phi$ $[10^{-2}]$ & $\mathrm{Re}_\mathrm{p}$ &
    \multicolumn{2}{c}{$\mu_\mathrm{r}$}\\
    \hline
    0.01  & 100  & full & 4 & $48.2\pm0.9$ & $0.14$ & $11.1$ & $0.3$\\
    0.1   & 100  & full & 4 & $48\pm3$   & $0.15$ & $8.9$ & $0.3$\\
    0.001 & 100  & full & 4 & $48.2\pm0.8$ & $0.13$ & $11.8$ & $0.3$\\
    0.01  & 50   & full & 4 & $48.2\pm0.7$ & $0.14$ & $10.9$ & $0.3$\\
    0.01  & 1000 & full & 4 & $48\pm1$   & $0.14$ & $10.9$ & $0.4$\\
    0.01  & 100  & normal only & 4 & $48\pm1$   & $0.16$ & $8.1$ & $0.2$\\
    0.04  & 100  & link-based  & 4 & $48.0\pm0.4$ & $0.10$ & $20$ & $1$\\[2ex]
    0.01  & 100  & full & 2 & $48\pm1$ & $0.15$ & $8.8$ & $0.4$\\
    0.01  & 100  & normal only & 2 & $48\pm3$ & $0.17$ & $5.9$ & $0.2$\\
    0.04  & 400  & link-based  & 2 & $46.4\pm0.5$ & $0.16$ & $9.9$ &$0.6$\\[2ex]
    0.01  & 100  & full & 8 & $48.4\pm0.7$ & $0.13$ & $12.4$ & $0.2$\\
    0.01  & 100  & normal only & 8 & $48.3\pm0.9$ & $0.14$ & $9.6$ & $0.2$\\
    0.04  & 25   & link-based  & 8 & $48.0\pm0.6$ & $0.08$ & $28$ & $1$\\
    \hline\hline
  \end{tabular}
  \caption{\label{tab:results:viscosity-spheres}Relative shear viscosity
    $\mu_\mathrm{r}$ of dense ($\Phi\approx0.5$) suspensions of spherical
    particles for different lubrication models, resolutions $R$, and
    short-range numerical parameters $h_\mathrm{c}$ and
    $\epsilon_\mathrm{c}$ ($h_\mathrm{c}^*$ and $\epsilon_\mathrm{c}^*$ in
    case of link-based lubrication corrections). The effective volume
    fraction $\Phi$ and particle Reynolds number
    $\mathrm{Re}_\mathrm{p}=4R^2\dot{\gamma}/\nu$ vary slightly due to the
    simulation setup.}
\end{table}

\tabref{results:viscosity-spheres} compares the viscosities resulting
from a variation of the short-range parameters for $\Phi=0.5$ where
the effect of lubrication interactions is strongest. Apparently,
varying $\epsilon_\mathrm{c}$ has no significant effect on the
resulting viscosities. Better resolving lubrication interactions by
reducing $h_\mathrm{c}$ to only $0.001$ leads to an increase of
$\mu_\mathrm{r}$ that is noticeable but still smaller than
$10\,\%$. Reducing the resolution of lubrication corrections, however,
by choosing $h_\mathrm{c}=0.1$ causes a reduction of the viscosity by
$20\,\%$ to a value that is actually closer to data without non-normal
corrections than to data from the full model at
$h_\mathrm{c}=0.01$. Variations in the volume fraction $\Phi$ or the
particle Reynolds number $\mathrm{Re}_\mathrm{p}$ appear too small to
be of significant influence here, the latter being in fact the
consequence of differing viscosities in the bulk of the
simulation. \tabref{results:viscosity-spheres} also shows the
viscosities computed for the different lubrication models at the
resolutions $R=2$ and $8$. For $R=2$, the viscosities are clearly
reduced with respect to $R=4$, in the case of the link-based model to
only $50\,\%$. Changing the resolution to $R=8$ increases the
viscosity by about $12\,\%$ for the full contact-based model while at
$R=2$ a reduction by $21\,\%$ is seen. If the Stokesian dynamics
results are assumed to represent the correct viscosities, both
increasing $R$ to $8$ and decreasing $h_\mathrm{c}$ to $0.001$
slightly improves the accuracy of the present model. From the only
small improvement induced by already considerable changes of $R$ and
$h_\mathrm{c}$ it can be concluded, however, that the viscosities
obtained with the present parameters are already close to the value
theoretically achieved for $h_\mathrm{c}=0$ and infinite
resolution. While doubling the spatial resolution in an LB simulation
at fixed $\tau$ increases the computational effort by a factor of
$2^5=32$, also reducing the short-range cut-off to
$h_\mathrm{c}=0.001$ moderately increases the computational cost by
demanding each LB time step to be subdivided in 20 instead of 10
sub-steps for the particle update in order to keep the simulation
stable. Compared to other three-dimensional LB suspension
models~\cite{hyvaluoma05,kromkamp06} at even higher resolution, the
present approach performs well in reproducing the viscosity already in
well-affordable simulations at $R=4$ and $h_\mathrm{c}=0.01$ thanks to
the inclusion of the non-normal lubrication corrections. Without
non-normal corrections the viscosity is closest to the Stokesian
dynamics results for the highest resolution $R=8$ and the differences
to data at $R=2$ and $R=4$ suggest that the numbers would converge at
even larger $R$. The same might be true for the link-based model but
here, between $R=4$ and $8$ an increase of still $40\,\%$ is visible
while $\mu_\mathrm{r}$ appears to be over-predicted already.

\begin{table}
  \centering
  \begin{tabular}{lrll@{\hspace*{0.8em}}ll@{\hspace*{0.8em}}r@{$\,\pm\,$}l}
    \hline\hline
    $h_\mathrm{c}^{(*)}$ & $\epsilon_\mathrm{c}^{(*)}$ & model &
    $R_\parallel$ & $\Phi$ $[10^{-2}]$ & $\mathrm{Re}_\mathrm{p}$ &
    \multicolumn{2}{c}{$\mu_\mathrm{r}$}\\
    \hline
    0.01  & 100  & full & 2 & $48.3\pm0.5$ & $0.18$ & $7.1$ & $0.2$\\
    0.1   & 100  & full & 2 & $48.3\pm0.5$ & $0.18$ & $7.0$ & $0.2$\\
    0.001 & 1000 & full & 2 & $48.4\pm0.4$ & $0.18$ & $7.0$ & $0.2$\\
    0.01  & 80   & full & 2 & $48.4\pm0.4$ & $0.18$ & $7.1$ & $0.2$\\
    0.01  & 1000 & full & 2 & $48.4\pm0.5$ & $0.18$ & $7.1$ & $0.3$\\
    0.01  & 100  & normal only & 2 & $48.3\pm0.4$ & $0.20$ & $4.9$ & $0.1$\\
    0.04  & 100  & link-based  & 2 & $48.4\pm0.5$ & $0.12$ & $18.8$&$0.9$\\[2ex]
    0.01  & 100  & full & 4 & $48.7\pm0.6$ & $0.16$ & $8.7$ & $0.2$\\
    0.01  & 100  & normal only & 4 & $48.7\pm0.6$ & $0.18$ & $6.2$ & $0.2$\\
    0.04  & 25   & link-based  & 4 & $48.7\pm0.5$ & $0.07$ & $38$ & $2$\\
    \hline\hline
  \end{tabular}
  \caption{\label{tab:results:viscosity-oblates}Data corresponding to
    \tabref{results:viscosity-spheres} for oblate spheroids with half axes
    $R_\parallel$ and $R_\perp=3R_\parallel$. $\mathrm{Re}_\mathrm{p}$ is computed
    based on an average radius $(R_\parallel R_\perp^2)^{1/3}$.}
\end{table}

\begin{table}
  \centering
  \begin{tabular}{lrlr@{\hspace*{0.8em}}ll@{\hspace*{0.8em}}r@{$\,\pm\,$}l}
    \hline\hline
    $h_\mathrm{c}^{(*)}$ & $\epsilon_\mathrm{c}^{(*)}$ & model &
    $R_\parallel$ & $\Phi$ $[10^{-2}]$ & $\mathrm{Re}_\mathrm{p}$ &
    \multicolumn{2}{c}{$\mu_\mathrm{r}$}\\
    \hline
    0.01  & 100  & full & 6  & $48.5\pm0.4$ & $0.08$ & $7.7$ & $0.1$\\
    0.1   & 100  & full & 6  & $48.4\pm0.4$ & $0.09$ & $6.9$ & $0.1$\\
    0.001 & 100  & full & 6  & $48.5\pm0.4$ & $0.08$ & $7.6$ & $0.1$\\
    0.01  & 50   & full & 6  & $48.5\pm0.4$ & $0.08$ & $7.6$ & $0.1$\\
    0.01  & 1000 & full & 6  & $48.5\pm0.3$ & $0.08$ & $7.63$ & $0.09$\\
    0.01  & 100  & normal only & 6  & $48.4\pm0.4$ & $0.10$ & $4.56$ & $0.06$\\
    0.04  & 100  & link-based  & 6  & $47.9\pm0.4$ & $0.07$ &$12.7$&$0.4$\\[2ex]
    0.01  & 100  & full & 12 & $49.3\pm0.5$ & $0.07$ & $9.7$ & $0.1$\\
    0.01  & 100  & normal only & 12 & $49.2\pm0.5$ & $0.09$ & $6.18$ & $0.09$\\
    0.04  & 25   & link-based  & 12 & $49.2\pm0.6$ & $0.04$ & $24.9$ & $0.6$\\
    \hline\hline
  \end{tabular}
  \caption{\label{tab:results:viscosity-prolates}Data corresponding to
    \tabref{results:viscosity-spheres} and \tabref{results:viscosity-oblates}
    for prolate spheroids with half axes $R_\parallel$ and
    $R_\perp=R_\parallel/3$.}
\end{table}

\picref{results:viscosity}(b) and (c) display the relative viscosity
of suspensions of spheroids as a function of the volume fraction for
the different lubrication models. In (b) the aspect ratio is
$\Lambda=1/3$ (oblates), in (c) it is $3$ (prolates). In both cases
the smaller half-axis is chosen to be $2$. As expected now, the
link-wise model leads to an enlarged $\mu_\mathrm{r}$, disabling
non-normal lubrication interactions in the contact-based model to a
reduced $\mu_\mathrm{r}$. For $\Phi\gtrsim0.3$ the full contact-based
model, which is believed to be most correct, predicts a clear
reduction of the viscosity of prolates as compared to spheres and of
oblates as compared to prolates. The result for oblates seems
inconsistent with the work of Bertevas \textit{et
  al.}~\cite{bertevas10} who report for oblate spheroids of aspect
ratio $0.3$ at volume fractions up to $\Phi=0.25$ a higher viscosity
than for spheres at the same volume fraction. Unfortunately,
Ref.~\cite{bertevas10} provides no data for spheroidal particles at
$\Phi>0.25$. One has to consider that in their work the non-singular
long-range hydrodynamic interactions of particles are neglected which
especially at lower volume fractions might cause particles to approach
closer than they would otherwise do. This would certainly increase the
viscosity. If, due to the large regions of low curvature and the
increased surface area this mechanism is stronger for oblates than for
spheres it could explain the inconsistency. Another possible
explanation is that in the model of Bertevas \textit{et
  al.}~\cite{bertevas10} the particle Reynolds number is strictly
$\mathrm{Re}_\mathrm{p}=0$ while in the present LB simulation it is
small but finite with $\mathrm{Re}_\mathrm{p}\sim10^{-1}$. It is hard
to explain, however, how inertia could cause a reduction of the
viscosity of oblates, especially since with respect to the viscosity
of sphere suspensions the method by Bertevas \textit{et
  al.}~\cite{bertevas10} hardly differs from the present LB model at
the respective volume fractions.

In \tabref{results:viscosity-oblates} and
\tabref{results:viscosity-prolates} the effect of varying the
short-range parameters and the resolution on the viscosity is
demonstrated at $\Phi=0.5$ for oblate and prolate spheroids. The
influence of the short-range parameters is smaller than for spheres
and only increasing $h_\mathrm{c}$ from $0.01$ to $0.1$ in the case of
prolates leads to a significant change in the suspension viscosity,
namely a reduction by about $10\,\%$. As for spheres, doubling the
resolution leads to higher viscosities. The increase is larger for
spheroids, in the case of the link-wise model it amounts to roughly
$100\,\%$ for oblates and prolates. The viscosities obtained from the
contact-based model, both with and without non-normal corrections,
increase by about $25\,\%$ and $35\,\%$ for oblates and prolates,
respectively. This result is in line with
\picref{results:spheres-club} and \picref{results:spheroid-resolution}
above that also demonstrate that due to their potentially smaller
local curvature, a higher resolution is required for spheroids than
for spheres in order to achieve a comparable degree of convergence. It
is interesting to examine the influence that the choice of lubrication
model has over the full range of volume fractions $\Phi$ in
\picref{results:viscosity} in the case of spheroids as compared to
spheres: while for spheres no significant influence is visible for
$\Phi\lesssim0.25$, the contact-based model without non-normal
interactions leads to a reduced viscosity for prolates already at
$\Phi\approx0.2$ and for oblates at the same volume fractions the
results of all three models differ. At larger $\Phi$ the discrepancy
between the models continues to grow. The finding is consistent with
the conclusion of Bertevas \textit{et al.}~\cite{bertevas10} stating
an increased importance of tangential lubrication interactions in
suspensions of oblate spheroids compared to spherical particles.

\section{Conclusions}\label{ch:conclusions}

The present paper implements a contact-based method for lubrication
corrections in the spirit of Nguyen and Ladd~\cite{nguyen02} that,
thanks to taking into account all leading singular terms of the
resistance matrix of the involved surfaces near
contact~\cite{cox74,claeys89}, is applicable also to spheroidal
particles. An extension to more general but smooth particle shapes is
straightforward as long as the contacts between particles are such
that they would touch in single points and an efficient method for
finding these contacts is known. An extension to shapes such as
cylinders, that might approach in line contacts, might be achievable
following the arguments of Butler and Shaqfeh~\cite{butler02}. Since
the lubrication corrections depend on the local properties of the
surfaces only, non-uniform particle dispersions with different sizes
and aspect ratios can be modeled in the present implementation
already.

In the case of spheres, the method shows high accuracy in the
resistances of two particles near contact as compared to theoretical
findings~\cite{jeffrey84}. The results obtained for self-diffusion and
viscosity in low Reynolds number shear flow of suspensions with volume
fractions between $\Phi=0.1$ and $0.5$ are highly consistent with
Stokesian dynamics simulations~\cite{sierou02,sierou04}. For these
results, a resolution of the sphere radius with $R=4$ lattice sites
suffices. For spheroids, the respective results appear consistent when
comparing simulations at different resolutions but due to the possibly
smaller local radii of curvature, discretization errors play a larger
role than for spheres and resolving the smaller half-axis with only
$1$ lattice unit cannot be recommended. For volume fractions
$\Phi\gtrsim0.3$, the model predicts a reduction of the suspension
viscosity of prolates with aspect ratio $\Lambda=3$ as compared to
spheres and a further reduction of oblates with aspect ratio $1/3$ as
compared to prolates.

Neglecting the non-normal lubrication corrections, as it is done by
many authors simulating spherical
particles~\cite{ladd97,hyvaluoma05,frijters12}, leads to an
under-estimation of shear-induced diffusion and viscosity. A link-wise
method for lubrication correction, similar to the initial work by Ding
and Aidun~\cite{ding03}, that over-predicts all non-normal lubrication
interactions, effects an over-estimation of shear-induced diffusion
and suspension viscosity. The error in the diffusion coefficient is
seen for all volume fractions $\Phi$. This is no surprise since it is
known that shear-induced diffusion can critically depend on the
short-range interactions of particles~\cite{dacunha96} and already
small inconsistencies summed up over many encounters can result in
considerable errors in the long-time behavior of the mean-square
displacement. Remarkably, the diffusion coefficients computed from the
full contact-based model at a resolution as low as $R=2$ appear more
consistent with diffusion coefficients from the same method at higher
resolution and from Stokesian dynamics simulations than the ones from
any of the other two methods tested at $R=8$. The viscosity of spheres
only slowly starts to depend on the non-normal corrections around
$\Phi=0.3$. For spheroids, however, an effect is found already around
$\Phi=0.2$. It can be concluded that while non-normal lubrication
corrections might indeed be of only minor importance for some
simulations involving spheres, their proper consideration is
essential, even at comparably large resolutions, such as $R=8$ for
spheres, once shear-induced diffusion of spherical or the viscosity of
spheroidal particles is of interest.

\begin{acknowledgments}
  The authors thank John F.~Brady and Francesco Picano for fruitful
  discussions. Financial support is gratefully acknowledged from the
  TU/e High Potential Research Program. Further, the authors
  acknowledge computing resources from JSC J\"ulich (through both GSC
  and PRACE grants), SSC Karlsruhe, and HLRS Stuttgart.
\end{acknowledgments}

\appendix*\label{ch:appendix}

\section{Diverging elements of $\mathbf{K}$}

The diverging elements of $\mathbf{K}$ that can be computed from the
local curvatures alone are directly taken from Cox~\cite{cox74} and
listed below only for completeness. The remaining elements that depend
also on $\Gamma_{0-3}$ and $\Gamma'_{0-3}$ and on the coefficients of
a fourth-order expansion can, in principle, be obtained by comparing
the negative forces and torques on the fluid as presented by Claeys
and Brady~\cite{claeys89} with \eqnref{lub:resistance}. However, it
has been noted~\cite{staben06} that the derivations of the remaining
diverging elements~\cite{claeys89} suffer from a sign error that
propagates into the equations (2.19a) to (2.20) of
Ref.~\cite{claeys89}. In consequence, the signs of (2.19c-d) are
flipped; (2.19a-b) and (2.20) are affected in a more complex
way. Following Staben \textit{et al.}~\cite{staben06} it is
straightforward to recalculate (2.19a-d) in the general case from
which \eqnref{app:chi13}, \eqnref{app:chi23}, \eqnref{app:chi34}, and
\eqnref{app:chi35} below are obtained. The results are verified by the
fact that following the same procedure while imposing the original
sign error~\cite{claeys89,staben06} yields exactly the original
terms~\cite{claeys89}. Assuming particle $j$ in
\eqnref{lub:resistance} to be a flat wall with velocity
$\mathbf{V}_j=\mathbf{0}$ lets \eqnref{app:chi34} produce a force or
torque identical to equations (2.12b) or (2.16a) in the work by Staben
\textit{et al.}~\cite{staben06}. Unfortunately, \eqnref{app:chi23}
results in just the opposite of (2.15b) or (2.16b) of
Ref.~\cite{staben06}. It is believed that this is caused either by a
typographical error or by an inconsistency in Ref.~\cite{staben06}
with respect to whether the equations describe the effect on the fluid
by the particle or vice versa. A recalculation of
(2.20)~\cite{claeys89} would be considerably more
involved~\cite{claeys89,staben06} and is therefore omitted which means
that the $\ln h$ contribution in \eqnref{app:chi33} is missing which
at small $h$, however, is dominated by the $h^{-1}$ term anyway.
\begin{widetext}
  \begin{equation}\label{eq:app:first-element}
    K_{11}
    =
    \frac
    {-\pi\ln h}
    {\sqrt{\lambda_1\lambda_2}S_1^2}
    \left[
      \frac
      {3\cos^2\chi(1-\lambda_1S_1)^2}
      {(3\lambda_1+2\lambda_2)\lambda_1}
      +
      \frac
      {3\sin^2\chi(1-\lambda_2S_1)^2}
      {(2\lambda_1+3\lambda_2)\lambda_2}
      +
      S_1^2
    \right]
  \end{equation}
  \begin{equation}
    K_{12}=K_{21}
    =
    \frac
    {3\pi\ln h}
    {\sqrt{\lambda_1\lambda_2}S_1S_2}
    \sin\chi
    \cos\chi
    \left[
      \frac
      {(1-\lambda_1S_1)(1-\lambda_1S_2)}
      {(3\lambda_1+2\lambda_2)\lambda_1}
      -
      \frac
      {(1-\lambda_2S_1)(1-\lambda_2S_2)}
      {(2\lambda_1+3\lambda_2)\lambda_2}
    \right]
  \end{equation}
  \begin{eqnarray}\label{eq:app:chi13}
    K_{13}=K_{31} &
    = &
    \frac
    {-3\pi\ln h}
    {2\sqrt{\lambda_1\lambda_2}(\lambda_1+\lambda_2)}\times
    \nonumber\\
    & &
    \Bigg[
    \left[
      2\sqrt{\lambda_1}(3\kappa_0\lambda_1+\kappa_2\lambda_2)
      -\frac{3}{2S_1\sqrt{\lambda_1}}
      ((7\lambda_1+2\lambda_2)\kappa_0+(\lambda_1+2\lambda_2)\kappa_2)
    \right]
    \frac{\cos\chi}{3\lambda_1+2\lambda_2}
    \nonumber\\
    & &{}+
    \left[
      2\sqrt{\lambda_2}(\kappa_1\lambda_1+3\kappa_3\lambda_2)
      -\frac{3}{2S_1\sqrt{\lambda_2}}
      ((2\lambda_1+\lambda_2)\kappa_1+(2\lambda_1+7\lambda_2)\kappa_3)
    \right]
    \frac{\sin\chi}{2\lambda_1+3\lambda_2}
    \nonumber\\
    & &{}+
    3\Gamma_0
    \left(
      \frac{\cos^2\chi}{\lambda_1}
      +
      \frac{\sin^2\chi}{\lambda_2}
    \right)
    +
    2\Gamma_1
    \sin\chi
    \cos\chi
    \left(
      \frac{1}{\lambda_2}
      -
      \frac{1}{\lambda_1}
    \right)
    +
    \Gamma_2
    \left(
      \frac{\sin^2\chi}{\lambda_1}
      +
      \frac{\cos^2\chi}{\lambda_2}
    \right)
    \Bigg]
  \end{eqnarray}
  \begin{equation}
    K_{14}=K_{41}
    =
    \frac
    {3\pi\ln h}
    {\sqrt{\lambda_1\lambda_2}S_1}
    \sin\chi
    \cos\chi
    \left[
      \frac
      {1-\lambda_1S_1}
      {(3\lambda_1+2\lambda_2)\lambda_1}
      -
      \frac
      {1-\lambda_2S_1}
      {(2\lambda_1+3\lambda_2)\lambda_2}
    \right]
  \end{equation}
  \begin{equation}
    K_{15}=K_{51}
    =
    \frac
    {3\pi\ln h}
    {\sqrt{\lambda_1\lambda_2}S_1}
    \left[
      \frac
      {\cos^2\chi(1-\lambda_1S_1)}
      {(3\lambda_1+2\lambda_2)\lambda_1}
      +
      \frac
      {\sin^2\chi(1-\lambda_2S_1)}
      {(2\lambda_1+3\lambda_2)\lambda_2}
    \right]
  \end{equation}
  \begin{equation}
    K_{22}
    =
    \frac
    {-\pi\ln h}
    {\sqrt{\lambda_1\lambda_2}S_2^2}
    \left[
      \frac
      {3\sin^2\chi(1-\lambda_1S_2)^2}
      {(3\lambda_1+2\lambda_2)\lambda_1}
      +
      \frac
      {3\cos^2\chi(1-\lambda_2S_2)^2}
      {(2\lambda_1+3\lambda_2)\lambda_2}
      +
      S_2^2
    \right]
  \end{equation}
  \begin{eqnarray}\label{eq:app:chi23}
    K_{23}=K_{32} &
    = &
    \frac
    {-3\pi\ln h}
    {2\sqrt{\lambda_1\lambda_2}(\lambda_1+\lambda_2)}\times
    \nonumber\\
    & &
    \Bigg[
    \left[
      -2\sqrt{\lambda_1}(3\kappa_0\lambda_1+\kappa_2\lambda_2)
      +\frac{3}{2S_2\sqrt{\lambda_1}}
      ((7\lambda_1+2\lambda_2)\kappa_0+(\lambda_1+2\lambda_2)\kappa_2)
    \right]
    \frac{\sin\chi}{3\lambda_1+2\lambda_2}
    \nonumber\\
    & &{}+
    \left[
      2\sqrt{\lambda_2}(\kappa_1\lambda_1+3\kappa_3\lambda_2)
      -\frac{3}{2S_2\sqrt{\lambda_2}}
      ((2\lambda_1+\lambda_2)\kappa_1+(2\lambda_1+7\lambda_2)\kappa_3)
    \right]
    \frac{\cos\chi}{2\lambda_1+3\lambda_2}
    \nonumber\\
    & &{}+
    3\Gamma_3
    \left(
      \frac{\sin^2\chi}{\lambda_1}
      +
      \frac{\cos^2\chi}{\lambda_2}
    \right)
    +
    2\Gamma_2\sin\chi\cos\chi
    \left(
      \frac{1}{\lambda_2}
      -
      \frac{1}{\lambda_1}
    \right)
    +
    \Gamma_1
    \left(
      \frac{\cos^2\chi}{\lambda_1}
      +
      \frac{\sin^2\chi}{\lambda_2}
    \right)
    \Bigg]
  \end{eqnarray}
  \begin{equation}
    K_{24}=K_{42}
    =
    \frac
    {-3\pi\ln h}
    {\sqrt{\lambda_1\lambda_2}S_2}
    \left[
      \frac
      {\sin^2\chi(1-\lambda_1S_2)}
      {(3\lambda_1+2\lambda_2)\lambda_1}
      +
      \frac
      {\cos^2\chi(1-\lambda_2S_2)}
      {(2\lambda_1+3\lambda_2)\lambda_2}
    \right]
  \end{equation}
  \begin{equation}
    K_{25}=K_{52}
    =
    \frac
    {3\pi\ln h}
    {\sqrt{\lambda_1\lambda_2}S_2}
    \sin\chi
    \cos\chi
    \left[
      -\frac
      {1-\lambda_1S_2}
      {(3\lambda_1+2\lambda_2)\lambda_1}
      +
      \frac
      {1-\lambda_2S_2}
      {(2\lambda_1+3\lambda_2)\lambda_2}
    \right]
  \end{equation}
  \begin{equation}\label{eq:app:chi33}
    K_{33}
    =
    \frac
    {3\pi}
    {h\sqrt{\lambda_1\lambda_2}(\lambda_1+\lambda_2)}
  \end{equation}
  \begin{equation}\label{eq:app:chi34}
    K_{34}=K_{43}
    =
    \frac
    {9\pi\ln h}
    {4\sqrt{\lambda_1\lambda_2}(\lambda_1+\lambda_2)}
    \left[
      \frac
      {(2\lambda_1+\lambda_2)\kappa_1 + (2\lambda_1+7\lambda_2)\kappa_3}
      {\sqrt{\lambda_2}(2\lambda_1+3\lambda_2)}
      \cos\chi
      -
      \frac
      {(\lambda_1+2\lambda_2)\kappa_2 + (7\lambda_1+2\lambda_2)\kappa_0}
      {\sqrt{\lambda_1}(3\lambda_1+2\lambda_2)}
      \sin\chi
    \right]
  \end{equation}
  \begin{equation}\label{eq:app:chi35}
    K_{35}=K_{53}
    =
    \frac
    {-9\pi\ln h}
    {4\sqrt{\lambda_1\lambda_2}(\lambda_1+\lambda_2)}
    \left[
      \frac
      {(\lambda_1+2\lambda_2)\kappa_2 + (7\lambda_1+2\lambda_2)\kappa_0}
      {\sqrt{\lambda_1}(3\lambda_1+2\lambda_2)}
      \cos\chi
      +
      \frac
      {(2\lambda_1+\lambda_2)\kappa_1 + (2\lambda_1+7\lambda_2)\kappa_3}
      {\sqrt{\lambda_2}(2\lambda_1+3\lambda_2)}
      \sin\chi
    \right]
  \end{equation}
  \begin{equation}
    K_{36}=K_{63}
    =
    \frac
    {-3\pi\ln h}
    {2\sqrt{\lambda_1\lambda_2}(\lambda_1+\lambda_2)}
    \sin\chi
    \cos\chi
    \left[
      \frac
      {1}
      {S_1}
      -
      \frac
      {1}
      {S_2}
    \right]
    \left[
      \frac
      {1}
      {\lambda_1}
      -
      \frac
      {1}
      {\lambda_2}
    \right]
  \end{equation}
  \begin{equation}
    K_{44}
    =
    \frac
    {-3\pi\ln h}
    {\sqrt{\lambda_1\lambda_2}}
    \left[
      \frac
      {\sin^2\chi}
      {(3\lambda_1+2\lambda_2)\lambda_1}
      +
      \frac
      {\cos^2\chi}
      {(2\lambda_1+3\lambda_2)\lambda_2}
    \right]
  \end{equation}
  \begin{equation}
    K_{45}=K_{54}
    =
    \frac
    {3\pi\ln h}
    {\sqrt{\lambda_1\lambda_2}}
    \sin\chi
    \cos\chi
    \left[
      \frac
      {-1}
      {(3\lambda_1+2\lambda_2)\lambda_1}
      +
      \frac
      {1}
      {(2\lambda_1+3\lambda_2)\lambda_2}
    \right]
  \end{equation}
  \begin{equation}\label{eq:app:last-element}
    K_{55}
    =
    \frac
    {-3\pi\ln{h}}
    {\sqrt{\lambda_1\lambda_2}}
    \left[
      \frac
      {\cos^2\chi}
      {(3\lambda_1+2\lambda_2)\lambda_1}
      +
      \frac
      {\sin^2\chi}
      {(2\lambda_1+3\lambda_2)\lambda_2}
    \right]
  \end{equation}
\end{widetext}
Some of the symbols in \eqnref{app:first-element} to
\eqnref{app:last-element} still require clarification. This
information is accessible also in Ref.~\cite{claeys89} and, partly,
\cite{cox74} but is summarized here to make the above description
self-contained. As visible in \picref{lub:cox74}, $\phi$ is the angle
between the axes of principal curvature $x_1$ and $x_1'$ and between
$x_2$ and $x_2'$ of both surfaces. In the derivation of the singular
terms, the height of the quadratically approximated gap between the
surfaces defined by \eqnref{lub:surface-i} and \eqnref{lub:surface-j}
is brought to the simple form
\begin{equation}
  h_z=1+\lambda_1\hat{x}_1^2+\lambda_2\hat{x}_2^2
\end{equation}
in terms of rescaled coordinates $\hat{x}_1$ and $\hat{x}_2$ where the
new principal curvatures $\lambda_1$ and $\lambda_2$ are the
eigenvalues of the matrix
\begin{equation*}
  \left(
    \begin{array}{l@{\quad}r}
      \displaystyle
      \frac{1}{2S_1}+\frac{\cos^2\phi}{2S_1'}+\frac{\sin^2\phi}{2S_2'} &
      \displaystyle
      \frac{\sin\phi\cos\phi}{2}\left(\frac{1}{S_1'}-
        \frac{1}{S_2'}\right)\\[3ex]
      \displaystyle
      \frac{\sin\phi\cos\phi}{2}\left(\frac{1}{S_1'}-\frac{1}{S_2'}\right) &
      \displaystyle
      \frac{1}{2S_2}+\frac{\sin^2\phi}{2S_1'}+\frac{\cos^2\phi}{2S_2'}
    \end{array}
  \right)
  \text{ .}
\end{equation*}
The trigonometric functions of the angle $\chi$ that transforms
between the directions of principal curvature $x_1$ and $x_2$ and the
main axes $\hat{x}_1$ and $\hat{x}_2$ can be obtained from the
components of the corresponding normalized eigenvectors
$\hat{\mathbf{l}}_1$ and $\hat{\mathbf{l}}_2$, which form the
transformation matrix, for instance
\begin{equation}
  \hat{\mathbf{l}}_2
  =
  \left(
    \begin{array}{c}
      \sin\chi\\
      \cos\chi
    \end{array}
  \right)
  \text{ .}
\end{equation}
The $\kappa_{0-3}$ describe the cubic features of both surfaces in the
coordinate frame defined by $\hat{x}_1$ and $\hat{x}_2$. Knowing
$\phi$, it is possible to express the cubic terms, characterized by
$\Gamma_{0-3}'$ in \eqnref{lub:surface-j}, in the principal frame
$x_1$ and $x_2$ of the other surface. The transformation can be
described by a set of $4$ functions $m_i(\alpha,a_0,a_1,a_2,a_3)$ of a
transformation angle $\alpha$ and a set of cubic coefficients
$a_{0-3}$ defined as
\begin{eqnarray}
  m_0 & = &
  a_0\cos^3\alpha - a_1\sin\alpha\cos^2\alpha\nonumber\\
  & &{}+ a_2\sin^2\alpha\cos\alpha - a_3\sin^3\alpha\nonumber\\
  m_1 & = &
  a_03\sin\alpha\cos^2\alpha +
  a_1(\cos^3\alpha-2\sin^2\alpha\cos\alpha)\nonumber\\
  & &{}+ a_2(\sin^3\alpha-2\sin\alpha\cos^2\alpha) +
  a_33\sin^2\alpha\cos\alpha\nonumber\\
  m_2 & = &
  a_03\cos\alpha\sin^2\alpha + 
  a_1(2\cos^2\alpha\sin\alpha-\sin^3\alpha)\nonumber\\
  & &{}+ a_2(\cos^3\alpha-2\cos\alpha\sin^2\alpha) - 
  a_3(3\cos^2\alpha\sin\alpha)\nonumber\\
  m_3 & = &
  a_0\sin^3\alpha + a_1\cos\alpha\sin^2\alpha\nonumber\\
  & &{}+ a_2\cos^2\alpha\sin\alpha + a_3\cos^3\alpha
\end{eqnarray}
and then reads
\begin{equation}
  \beta_i
  =
  m_i(\phi,\Gamma_0',\Gamma_1',\Gamma_2',\Gamma_3')
  \text{ .}
\end{equation}
The same functional dependency is used to transform the added cubic
coefficients $k_i=\Gamma_i+\beta_i$ into the frame of $\hat{x}_1$ and
$\hat{x}_2$. The rescaled coordinates demand rescaling also of the
cubic coefficients to obtain
\begin{equation}
  \kappa_i
  =
  \frac
  {m_i(\chi,k_0,k_1,k_2,k_3)}
  {\sqrt{\lambda_1}^{3-i}\sqrt{\lambda_2}^i}
  \text{ .}
\end{equation}

\begin{thebibliography}{50}%
\makeatletter
\providecommand \@ifxundefined [1]{%
 \@ifx{#1\undefined}
}%
\providecommand \@ifnum [1]{%
 \ifnum #1\expandafter \@firstoftwo
 \else \expandafter \@secondoftwo
 \fi
}%
\providecommand \@ifx [1]{%
 \ifx #1\expandafter \@firstoftwo
 \else \expandafter \@secondoftwo
 \fi
}%
\providecommand \natexlab [1]{#1}%
\providecommand \enquote  [1]{``#1''}%
\providecommand \bibnamefont  [1]{#1}%
\providecommand \bibfnamefont [1]{#1}%
\providecommand \citenamefont [1]{#1}%
\providecommand \href@noop [0]{\@secondoftwo}%
\providecommand \href [0]{\begingroup \@sanitize@url \@href}%
\providecommand \@href[1]{\@@startlink{#1}\@@href}%
\providecommand \@@href[1]{\endgroup#1\@@endlink}%
\providecommand \@sanitize@url [0]{\catcode `\\12\catcode `\$12\catcode
  `\&12\catcode `\#12\catcode `\^12\catcode `\_12\catcode `\%12\relax}%
\providecommand \@@startlink[1]{}%
\providecommand \@@endlink[0]{}%
\providecommand \url  [0]{\begingroup\@sanitize@url \@url }%
\providecommand \@url [1]{\endgroup\@href {#1}{\urlprefix }}%
\providecommand \urlprefix  [0]{URL }%
\providecommand \Eprint [0]{\href }%
\providecommand \doibase [0]{http://dx.doi.org/}%
\providecommand \selectlanguage [0]{\@gobble}%
\providecommand \bibinfo  [0]{\@secondoftwo}%
\providecommand \bibfield  [0]{\@secondoftwo}%
\providecommand \translation [1]{[#1]}%
\providecommand \BibitemOpen [0]{}%
\providecommand \bibitemStop [0]{}%
\providecommand \bibitemNoStop [0]{.\EOS\space}%
\providecommand \EOS [0]{\spacefactor3000\relax}%
\providecommand \BibitemShut  [1]{\csname bibitem#1\endcsname}%
\let\auto@bib@innerbib\@empty
\bibitem [{\citenamefont {Lopez}\ and\ \citenamefont {Graham}(2007)}]{lopez07}%
  \BibitemOpen
  \bibfield  {author} {\bibinfo {author} {\bibfnamefont {M.}~\bibnamefont
  {Lopez}}\ and\ \bibinfo {author} {\bibfnamefont {M.~D.}\ \bibnamefont
  {Graham}},\ }\href@noop {} {\bibfield  {journal} {\bibinfo  {journal}
  {Physics of Fluids}\ }\textbf {\bibinfo {volume} {19}},\ \bibinfo {pages}
  {073602} (\bibinfo {year} {2007})}\BibitemShut {NoStop}%
\bibitem [{\citenamefont {Clausen}\ \emph {et~al.}(2011)\citenamefont
  {Clausen}, \citenamefont {Reasor},\ and\ \citenamefont {Aidun}}]{clausen11}%
  \BibitemOpen
  \bibfield  {author} {\bibinfo {author} {\bibfnamefont {J.~R.}\ \bibnamefont
  {Clausen}}, \bibinfo {author} {\bibfnamefont {D.~A.}\ \bibnamefont {Reasor}},
  \ and\ \bibinfo {author} {\bibfnamefont {C.~K.}\ \bibnamefont {Aidun}},\
  }\href@noop {} {\bibfield  {journal} {\bibinfo  {journal} {Journal of Fluid
  Mechanics}\ }\textbf {\bibinfo {volume} {685}},\ \bibinfo {pages} {202}
  (\bibinfo {year} {2011})}\BibitemShut {NoStop}%
\bibitem [{\citenamefont {Kr{\"u}ger}(2012)}]{kruger12}%
  \BibitemOpen
  \bibfield  {author} {\bibinfo {author} {\bibfnamefont {T.}~\bibnamefont
  {Kr{\"u}ger}},\ }\emph {\bibinfo {title} {Computer simulation study of
  collective phenomena in dense suspensions of red blood cells under shear}},\
  \href@noop {} {Ph.D. thesis},\ \bibinfo  {school} {Ruhr-Universit\"at Bochum}
  (\bibinfo {year} {2012})\BibitemShut {NoStop}%
\bibitem [{\citenamefont {Zhao}\ \emph {et~al.}(2012)\citenamefont {Zhao},
  \citenamefont {Shaqfeh},\ and\ \citenamefont {Narsimhan}}]{zhao12}%
  \BibitemOpen
  \bibfield  {author} {\bibinfo {author} {\bibfnamefont {H.}~\bibnamefont
  {Zhao}}, \bibinfo {author} {\bibfnamefont {E.~S.~G.}\ \bibnamefont
  {Shaqfeh}}, \ and\ \bibinfo {author} {\bibfnamefont {V.}~\bibnamefont
  {Narsimhan}},\ }\href@noop {} {\bibfield  {journal} {\bibinfo  {journal}
  {Physics of Fluids}\ }\textbf {\bibinfo {volume} {24}},\ \bibinfo {pages}
  {011902} (\bibinfo {year} {2012})}\BibitemShut {NoStop}%
\bibitem [{\citenamefont {Grandchamp}\ \emph {et~al.}(2013)\citenamefont
  {Grandchamp}, \citenamefont {Coupier}, \citenamefont {Srivastav},
  \citenamefont {Minetti},\ and\ \citenamefont {Podgorski}}]{grandchamp13}%
  \BibitemOpen
  \bibfield  {author} {\bibinfo {author} {\bibfnamefont {X.}~\bibnamefont
  {Grandchamp}}, \bibinfo {author} {\bibfnamefont {G.}~\bibnamefont {Coupier}},
  \bibinfo {author} {\bibfnamefont {A.}~\bibnamefont {Srivastav}}, \bibinfo
  {author} {\bibfnamefont {C.}~\bibnamefont {Minetti}}, \ and\ \bibinfo
  {author} {\bibfnamefont {T.}~\bibnamefont {Podgorski}},\ }\href@noop {}
  {\bibfield  {journal} {\bibinfo  {journal} {{Phys. Rev. Lett.}}\ }\textbf
  {\bibinfo {volume} {110}},\ \bibinfo {pages} {108101} (\bibinfo {year}
  {2013})}\BibitemShut {NoStop}%
\bibitem [{\citenamefont {Omori}\ \emph {et~al.}(2013)\citenamefont {Omori},
  \citenamefont {Ishikawa}, \citenamefont {Imai},\ and\ \citenamefont
  {Yamaguchi}}]{omori13}%
  \BibitemOpen
  \bibfield  {author} {\bibinfo {author} {\bibfnamefont {T.}~\bibnamefont
  {Omori}}, \bibinfo {author} {\bibfnamefont {T.}~\bibnamefont {Ishikawa}},
  \bibinfo {author} {\bibfnamefont {Y.}~\bibnamefont {Imai}}, \ and\ \bibinfo
  {author} {\bibfnamefont {T.}~\bibnamefont {Yamaguchi}},\ }\href@noop {}
  {\bibfield  {journal} {\bibinfo  {journal} {{J. Fluid Mech.}}\ }\textbf
  {\bibinfo {volume} {724}},\ \bibinfo {pages} {154} (\bibinfo {year}
  {2013})}\BibitemShut {NoStop}%
\bibitem [{\citenamefont {Yeo}\ and\ \citenamefont {Maxey}(2013)}]{yeo13}%
  \BibitemOpen
  \bibfield  {author} {\bibinfo {author} {\bibfnamefont {K.}~\bibnamefont
  {Yeo}}\ and\ \bibinfo {author} {\bibfnamefont {M.~R.}\ \bibnamefont
  {Maxey}},\ }\href {\doibase 10.1063/1.4802844} {\bibfield  {journal}
  {\bibinfo  {journal} {Physics of Fluids}\ }\textbf {\bibinfo {volume} {25}},\
  \bibinfo {eid} {053303} (\bibinfo {year} {2013})}\BibitemShut {NoStop}%
\bibitem [{\citenamefont {Metzger}\ \emph {et~al.}(2013)\citenamefont
  {Metzger}, \citenamefont {Pham},\ and\ \citenamefont {Butler}}]{metzger13}%
  \BibitemOpen
  \bibfield  {author} {\bibinfo {author} {\bibfnamefont {B.}~\bibnamefont
  {Metzger}}, \bibinfo {author} {\bibfnamefont {P.}~\bibnamefont {Pham}}, \
  and\ \bibinfo {author} {\bibfnamefont {J.~E.}\ \bibnamefont {Butler}},\
  }\href {\doibase 10.1103/PhysRevE.87.052304} {\bibfield  {journal} {\bibinfo
  {journal} {Phys. Rev. E}\ }\textbf {\bibinfo {volume} {87}},\ \bibinfo
  {pages} {052304} (\bibinfo {year} {2013})}\BibitemShut {NoStop}%
\bibitem [{\citenamefont {Goldman}\ \emph {et~al.}(1967)\citenamefont
  {Goldman}, \citenamefont {Cox},\ and\ \citenamefont {Brenner}}]{goldman67}%
  \BibitemOpen
  \bibfield  {author} {\bibinfo {author} {\bibfnamefont {A.}~\bibnamefont
  {Goldman}}, \bibinfo {author} {\bibfnamefont {R.}~\bibnamefont {Cox}}, \ and\
  \bibinfo {author} {\bibfnamefont {H.}~\bibnamefont {Brenner}},\ }\href
  {\doibase 10.1016/0009-2509(67)80047-2} {\bibfield  {journal} {\bibinfo
  {journal} {Chemical Engineering Science}\ }\textbf {\bibinfo {volume} {22}},\
  \bibinfo {pages} {637 } (\bibinfo {year} {1967})}\BibitemShut {NoStop}%
\bibitem [{\citenamefont {Cox}(1974)}]{cox74}%
  \BibitemOpen
  \bibfield  {author} {\bibinfo {author} {\bibfnamefont {R.~G.}\ \bibnamefont
  {Cox}},\ }\href {\doibase 10.1016/0301-9322(74)90019-6} {\bibfield  {journal}
  {\bibinfo  {journal} {International Journal of Multiphase Flow}\ }\textbf
  {\bibinfo {volume} {1}},\ \bibinfo {pages} {343} (\bibinfo {year}
  {1974})}\BibitemShut {NoStop}%
\bibitem [{\citenamefont {Claeys}\ and\ \citenamefont
  {Brady}(1989)}]{claeys89}%
  \BibitemOpen
  \bibfield  {author} {\bibinfo {author} {\bibfnamefont {I.~L.}\ \bibnamefont
  {Claeys}}\ and\ \bibinfo {author} {\bibfnamefont {J.~F.}\ \bibnamefont
  {Brady}},\ }\href@noop {} {\bibfield  {journal} {\bibinfo  {journal}
  {PhysicoChem. Hydrodyn}\ }\textbf {\bibinfo {volume} {11}},\ \bibinfo {pages}
  {261} (\bibinfo {year} {1989})}\BibitemShut {NoStop}%
\bibitem [{\citenamefont {Claeys}\ and\ \citenamefont
  {Brady}(1993)}]{claeys93}%
  \BibitemOpen
  \bibfield  {author} {\bibinfo {author} {\bibfnamefont {I.~L.}\ \bibnamefont
  {Claeys}}\ and\ \bibinfo {author} {\bibfnamefont {J.~F.}\ \bibnamefont
  {Brady}},\ }\href@noop {} {\bibfield  {journal} {\bibinfo  {journal} {{J.
  Fluid Mech.}}\ }\textbf {\bibinfo {volume} {251}},\ \bibinfo {pages} {411}
  (\bibinfo {year} {1993})}\BibitemShut {NoStop}%
\bibitem [{\citenamefont {Bertevas}\ \emph {et~al.}(2010)\citenamefont
  {Bertevas}, \citenamefont {Fan},\ and\ \citenamefont {Tanner}}]{bertevas10}%
  \BibitemOpen
  \bibfield  {author} {\bibinfo {author} {\bibfnamefont {E.}~\bibnamefont
  {Bertevas}}, \bibinfo {author} {\bibfnamefont {X.}~\bibnamefont {Fan}}, \
  and\ \bibinfo {author} {\bibfnamefont {R.~I.}\ \bibnamefont {Tanner}},\
  }\href {\doibase 10.1007/s00397-009-0390-8} {\bibfield  {journal} {\bibinfo
  {journal} {{Rheol. Acta}}\ }\textbf {\bibinfo {volume} {49}},\ \bibinfo
  {pages} {53} (\bibinfo {year} {2010})}\BibitemShut {NoStop}%
\bibitem [{\citenamefont {Succi}(2001)}]{succi01}%
  \BibitemOpen
  \bibfield  {author} {\bibinfo {author} {\bibfnamefont {S.}~\bibnamefont
  {Succi}},\ }\href@noop {} {\emph {\bibinfo {title} {{The Lattice {B}oltzmann
  Equation for Fluid Dynamics and Beyond}}}},\ {Numerical Mathematics and
  Scientific Computation}\ (\bibinfo  {publisher} {{Oxford University Press}},\
  \bibinfo {year} {2001})\BibitemShut {NoStop}%
\bibitem [{\citenamefont {Ladd}(1994{\natexlab{a}})}]{ladd94}%
  \BibitemOpen
  \bibfield  {author} {\bibinfo {author} {\bibfnamefont {A.~J.~C.}\
  \bibnamefont {Ladd}},\ }\href {\doibase 10.1017/S0022112094001771} {\bibfield
   {journal} {\bibinfo  {journal} {{J. Fluid Mech.}}\ }\textbf {\bibinfo
  {volume} {271}},\ \bibinfo {pages} {285} (\bibinfo {year}
  {1994}{\natexlab{a}})}\BibitemShut {NoStop}%
\bibitem [{\citenamefont {Ladd}(1994{\natexlab{b}})}]{ladd94b}%
  \BibitemOpen
  \bibfield  {author} {\bibinfo {author} {\bibfnamefont {A.~J.~C.}\
  \bibnamefont {Ladd}},\ }\href {\doibase 10.1017/S0022112094001783} {\bibfield
   {journal} {\bibinfo  {journal} {{J. Fluid Mech.}}\ }\textbf {\bibinfo
  {volume} {271}},\ \bibinfo {pages} {311} (\bibinfo {year}
  {1994}{\natexlab{b}})}\BibitemShut {NoStop}%
\bibitem [{\citenamefont {Ladd}\ and\ \citenamefont {Verberg}(2001)}]{ladd01}%
  \BibitemOpen
  \bibfield  {author} {\bibinfo {author} {\bibfnamefont {A.~J.~C.}\
  \bibnamefont {Ladd}}\ and\ \bibinfo {author} {\bibfnamefont {R.}~\bibnamefont
  {Verberg}},\ }\href {\doibase 10.1023/A:1010414013942} {\bibfield  {journal}
  {\bibinfo  {journal} {{J. Stat. Phys.}}\ }\textbf {\bibinfo {volume} {104}},\
  \bibinfo {pages} {1191} (\bibinfo {year} {2001})}\BibitemShut {NoStop}%
\bibitem [{\citenamefont {Aidun}\ and\ \citenamefont
  {Clausen}(2010)}]{aidun10}%
  \BibitemOpen
  \bibfield  {author} {\bibinfo {author} {\bibfnamefont {C.~K.}\ \bibnamefont
  {Aidun}}\ and\ \bibinfo {author} {\bibfnamefont {J.~R.}\ \bibnamefont
  {Clausen}},\ }\href {\doibase 10.1146/annurev-fluid-121108-145519} {\bibfield
   {journal} {\bibinfo  {journal} {{Annu. Rev. Fluid Mech.}}\ }\textbf
  {\bibinfo {volume} {42}},\ \bibinfo {pages} {439} (\bibinfo {year}
  {2010})}\BibitemShut {NoStop}%
\bibitem [{\citenamefont {Kunert}\ \emph {et~al.}(2010)\citenamefont {Kunert},
  \citenamefont {Harting},\ and\ \citenamefont {Vinogradova}}]{kunert10}%
  \BibitemOpen
  \bibfield  {author} {\bibinfo {author} {\bibfnamefont {C.}~\bibnamefont
  {Kunert}}, \bibinfo {author} {\bibfnamefont {J.}~\bibnamefont {Harting}}, \
  and\ \bibinfo {author} {\bibfnamefont {O.~I.}\ \bibnamefont {Vinogradova}},\
  }\href {\doibase 10.1103/PhysRevLett.105.016001} {\bibfield  {journal}
  {\bibinfo  {journal} {{Phys. Rev. Lett.}}\ }\textbf {\bibinfo {volume}
  {105}},\ \bibinfo {pages} {016001} (\bibinfo {year} {2010})}\BibitemShut
  {NoStop}%
\bibitem [{\citenamefont {Kunert}\ and\ \citenamefont
  {Harting}(2011)}]{kunert11}%
  \BibitemOpen
  \bibfield  {author} {\bibinfo {author} {\bibfnamefont {C.}~\bibnamefont
  {Kunert}}\ and\ \bibinfo {author} {\bibfnamefont {J.}~\bibnamefont
  {Harting}},\ }\href {\doibase 10.1093/imamat/hxr001} {\bibfield  {journal}
  {\bibinfo  {journal} {IMA Journal of Applied Mathematics}\ }\textbf {\bibinfo
  {volume} {76}},\ \bibinfo {pages} {761} (\bibinfo {year} {2011})}\BibitemShut
  {NoStop}%
\bibitem [{\citenamefont {Schwarzer}(1995)}]{schwarzer95}%
  \BibitemOpen
  \bibfield  {author} {\bibinfo {author} {\bibfnamefont {S.}~\bibnamefont
  {Schwarzer}},\ }\href@noop {} {\bibfield  {journal} {\bibinfo  {journal}
  {{Phys. Rev. E}}\ }\textbf {\bibinfo {volume} {52}},\ \bibinfo {pages} {6461}
  (\bibinfo {year} {1995})}\BibitemShut {NoStop}%
\bibitem [{\citenamefont {Hecht}\ \emph {et~al.}(2005)\citenamefont {Hecht},
  \citenamefont {Harting}, \citenamefont {Ihle},\ and\ \citenamefont
  {Herrmann}}]{hecht05}%
  \BibitemOpen
  \bibfield  {author} {\bibinfo {author} {\bibfnamefont {M.}~\bibnamefont
  {Hecht}}, \bibinfo {author} {\bibfnamefont {J.}~\bibnamefont {Harting}},
  \bibinfo {author} {\bibfnamefont {T.}~\bibnamefont {Ihle}}, \ and\ \bibinfo
  {author} {\bibfnamefont {H.~J.}\ \bibnamefont {Herrmann}},\ }\href {\doibase
  10.1103/PhysRevE.72.011408} {\bibfield  {journal} {\bibinfo  {journal}
  {{Phys. Rev. E}}\ }\textbf {\bibinfo {volume} {72}},\ \bibinfo {pages}
  {011408} (\bibinfo {year} {2005})}\BibitemShut {NoStop}%
\bibitem [{\citenamefont {Martys}(2005)}]{martys05}%
  \BibitemOpen
  \bibfield  {author} {\bibinfo {author} {\bibfnamefont {N.~S.}\ \bibnamefont
  {Martys}},\ }\href {\doibase 10.1122/1.1849187} {\bibfield  {journal}
  {\bibinfo  {journal} {Journal of Rheology}\ }\textbf {\bibinfo {volume}
  {49}},\ \bibinfo {pages} {401} (\bibinfo {year} {2005})}\BibitemShut
  {NoStop}%
\bibitem [{\citenamefont {Ladd}(1997)}]{ladd97}%
  \BibitemOpen
  \bibfield  {author} {\bibinfo {author} {\bibfnamefont {A.~J.~C.}\
  \bibnamefont {Ladd}},\ }\href {\doibase 10.1063/1.869212} {\bibfield
  {journal} {\bibinfo  {journal} {Physics of Fluids}\ }\textbf {\bibinfo
  {volume} {9}},\ \bibinfo {pages} {491} (\bibinfo {year} {1997})}\BibitemShut
  {NoStop}%
\bibitem [{\citenamefont {Hyväluoma}\ \emph {et~al.}(2005)\citenamefont
  {Hyväluoma}, \citenamefont {Raiskinmäki}, \citenamefont {Koponen},
  \citenamefont {Kataja},\ and\ \citenamefont {Timonen}}]{hyvaluoma05}%
  \BibitemOpen
  \bibfield  {author} {\bibinfo {author} {\bibfnamefont {J.}~\bibnamefont
  {Hyväluoma}}, \bibinfo {author} {\bibfnamefont {P.}~\bibnamefont
  {Raiskinmäki}}, \bibinfo {author} {\bibfnamefont {A.}~\bibnamefont
  {Koponen}}, \bibinfo {author} {\bibfnamefont {M.}~\bibnamefont {Kataja}}, \
  and\ \bibinfo {author} {\bibfnamefont {J.}~\bibnamefont {Timonen}},\ }\href
  {\doibase 10.1007/s10955-005-4314-4} {\bibfield  {journal} {\bibinfo
  {journal} {{J. Stat. Phys.}}\ }\textbf {\bibinfo {volume} {121}},\ \bibinfo
  {pages} {149} (\bibinfo {year} {2005})}\BibitemShut {NoStop}%
\bibitem [{\citenamefont {Kromkamp}\ \emph {et~al.}(2006)\citenamefont
  {Kromkamp}, \citenamefont {van~den Ende}, \citenamefont {Kandhai},
  \citenamefont {van~der Sman},\ and\ \citenamefont {Boom}}]{kromkamp06}%
  \BibitemOpen
  \bibfield  {author} {\bibinfo {author} {\bibfnamefont {J.}~\bibnamefont
  {Kromkamp}}, \bibinfo {author} {\bibfnamefont {D.~T.~M.}\ \bibnamefont
  {van~den Ende}}, \bibinfo {author} {\bibfnamefont {D.}~\bibnamefont
  {Kandhai}}, \bibinfo {author} {\bibfnamefont {R.~G.~M.}\ \bibnamefont
  {van~der Sman}}, \ and\ \bibinfo {author} {\bibfnamefont {R.~M.}\
  \bibnamefont {Boom}},\ }\href@noop {} {\bibfield  {journal} {\bibinfo
  {journal} {Chemical engineering science}\ }\textbf {\bibinfo {volume} {61}},\
  \bibinfo {pages} {858} (\bibinfo {year} {2006})}\BibitemShut {NoStop}%
\bibitem [{\citenamefont {Nguyen}\ and\ \citenamefont {Ladd}(2002)}]{nguyen02}%
  \BibitemOpen
  \bibfield  {author} {\bibinfo {author} {\bibfnamefont {N.-Q.}\ \bibnamefont
  {Nguyen}}\ and\ \bibinfo {author} {\bibfnamefont {A.~J.~C.}\ \bibnamefont
  {Ladd}},\ }\href {\doibase 10.1103/PhysRevE.66.046708} {\bibfield  {journal}
  {\bibinfo  {journal} {{Phys. Rev. E}}\ }\textbf {\bibinfo {volume} {66}},\
  \bibinfo {pages} {046708} (\bibinfo {year} {2002})}\BibitemShut {NoStop}%
\bibitem [{\citenamefont {Qi}\ \emph {et~al.}(2002)\citenamefont {Qi},
  \citenamefont {Luo}, \citenamefont {Aravamuthan},\ and\ \citenamefont
  {Strieder}}]{qi02}%
  \BibitemOpen
  \bibfield  {author} {\bibinfo {author} {\bibfnamefont {D.}~\bibnamefont
  {Qi}}, \bibinfo {author} {\bibfnamefont {L.}~\bibnamefont {Luo}}, \bibinfo
  {author} {\bibfnamefont {R.}~\bibnamefont {Aravamuthan}}, \ and\ \bibinfo
  {author} {\bibfnamefont {W.}~\bibnamefont {Strieder}},\ }\href {\doibase
  10.1023/A:1014502402884} {\bibfield  {journal} {\bibinfo  {journal} {{J.
  Stat. Phys.}}\ }\textbf {\bibinfo {volume} {107}},\ \bibinfo {pages} {101}
  (\bibinfo {year} {2002})}\BibitemShut {NoStop}%
\bibitem [{\citenamefont {Günther}\ \emph {et~al.}(2013)\citenamefont
  {Günther}, \citenamefont {Janoschek}, \citenamefont {Frijters},\ and\
  \citenamefont {Harting}}]{gunther13}%
  \BibitemOpen
  \bibfield  {author} {\bibinfo {author} {\bibfnamefont {F.}~\bibnamefont
  {Günther}}, \bibinfo {author} {\bibfnamefont {F.}~\bibnamefont {Janoschek}},
  \bibinfo {author} {\bibfnamefont {S.}~\bibnamefont {Frijters}}, \ and\
  \bibinfo {author} {\bibfnamefont {J.}~\bibnamefont {Harting}},\ }\href
  {\doibase 10.1016/j.compfluid.2012.03.020} {\bibfield  {journal} {\bibinfo
  {journal} {Computers \& Fluids}\ }\textbf {\bibinfo {volume} {80}},\ \bibinfo
  {pages} {184 } (\bibinfo {year} {2013})}\BibitemShut {NoStop}%
\bibitem [{\citenamefont {Janoschek}\ \emph {et~al.}(2010)\citenamefont
  {Janoschek}, \citenamefont {Toschi},\ and\ \citenamefont
  {Harting}}]{janoschek10}%
  \BibitemOpen
  \bibfield  {author} {\bibinfo {author} {\bibfnamefont {F.}~\bibnamefont
  {Janoschek}}, \bibinfo {author} {\bibfnamefont {F.}~\bibnamefont {Toschi}}, \
  and\ \bibinfo {author} {\bibfnamefont {J.}~\bibnamefont {Harting}},\ }\href
  {\doibase 10.1103/PhysRevE.82.056710} {\bibfield  {journal} {\bibinfo
  {journal} {{Phys. Rev. E}}\ }\textbf {\bibinfo {volume} {82}},\ \bibinfo
  {pages} {056710} (\bibinfo {year} {2010})}\BibitemShut {NoStop}%
\bibitem [{\citenamefont {Ding}\ and\ \citenamefont {Aidun}(2003)}]{ding03}%
  \BibitemOpen
  \bibfield  {author} {\bibinfo {author} {\bibfnamefont {E.-J.}\ \bibnamefont
  {Ding}}\ and\ \bibinfo {author} {\bibfnamefont {C.~K.}\ \bibnamefont
  {Aidun}},\ }\href {\doibase 10.1023/A:1023880126272} {\bibfield  {journal}
  {\bibinfo  {journal} {{J. Stat. Phys.}}\ }\textbf {\bibinfo {volume} {112}},\
  \bibinfo {pages} {685} (\bibinfo {year} {2003})}\BibitemShut {NoStop}%
\bibitem [{\citenamefont {MacMeccan}\ \emph {et~al.}(2009)\citenamefont
  {MacMeccan}, \citenamefont {Clausen}, \citenamefont {Neitzel},\ and\
  \citenamefont {Aidun}}]{macmeccan09}%
  \BibitemOpen
  \bibfield  {author} {\bibinfo {author} {\bibfnamefont {R.~M.}\ \bibnamefont
  {MacMeccan}}, \bibinfo {author} {\bibfnamefont {J.~R.}\ \bibnamefont
  {Clausen}}, \bibinfo {author} {\bibfnamefont {G.~P.}\ \bibnamefont
  {Neitzel}}, \ and\ \bibinfo {author} {\bibfnamefont {C.~K.}\ \bibnamefont
  {Aidun}},\ }\href@noop {} {\bibfield  {journal} {\bibinfo  {journal} {Journal
  of Fluid Mechanics}\ }\textbf {\bibinfo {volume} {618}},\ \bibinfo {pages}
  {13} (\bibinfo {year} {2009})}\BibitemShut {NoStop}%
\bibitem [{\citenamefont {Clausen}(2010)}]{clausen10b}%
  \BibitemOpen
  \bibfield  {author} {\bibinfo {author} {\bibfnamefont {J.~R.}\ \bibnamefont
  {Clausen}},\ }\emph {\bibinfo {title} {The effect of particle deformation on
  the rheology and microstructure of noncolloidal suspensions}},\ \href@noop {}
  {Ph.D. thesis},\ \bibinfo  {school} {Georgia Institute of Technology}
  (\bibinfo {year} {2010})\BibitemShut {NoStop}%
\bibitem [{\citenamefont {Qian}\ \emph {et~al.}(1992)\citenamefont {Qian},
  \citenamefont {{d'Humi\`eres}},\ and\ \citenamefont {Lallemand}}]{qian92}%
  \BibitemOpen
  \bibfield  {author} {\bibinfo {author} {\bibfnamefont {Y.~H.}\ \bibnamefont
  {Qian}}, \bibinfo {author} {\bibfnamefont {D.}~\bibnamefont
  {{d'Humi\`eres}}}, \ and\ \bibinfo {author} {\bibfnamefont {P.}~\bibnamefont
  {Lallemand}},\ }\href {\doibase 10.1209/0295-5075/17/6/001} {\bibfield
  {journal} {\bibinfo  {journal} {{Europhys. Lett.}}\ }\textbf {\bibinfo
  {volume} {17}},\ \bibinfo {pages} {479} (\bibinfo {year} {1992})}\BibitemShut
  {NoStop}%
\bibitem [{\citenamefont {Aidun}\ \emph {et~al.}(1998)\citenamefont {Aidun},
  \citenamefont {Lu},\ and\ \citenamefont {Ding}}]{aidun98}%
  \BibitemOpen
  \bibfield  {author} {\bibinfo {author} {\bibfnamefont {C.~K.}\ \bibnamefont
  {Aidun}}, \bibinfo {author} {\bibfnamefont {Y.}~\bibnamefont {Lu}}, \ and\
  \bibinfo {author} {\bibfnamefont {E.-J.}\ \bibnamefont {Ding}},\ }\href@noop
  {} {\bibfield  {journal} {\bibinfo  {journal} {{J. Fluid Mech.}}\ }\textbf
  {\bibinfo {volume} {373}},\ \bibinfo {pages} {287} (\bibinfo {year}
  {1998})}\BibitemShut {NoStop}%
\bibitem [{\citenamefont {Janoschek}\ \emph {et~al.}(2011)\citenamefont
  {Janoschek}, \citenamefont {Toschi},\ and\ \citenamefont
  {Harting}}]{janoschek11b}%
  \BibitemOpen
  \bibfield  {author} {\bibinfo {author} {\bibfnamefont {F.}~\bibnamefont
  {Janoschek}}, \bibinfo {author} {\bibfnamefont {F.}~\bibnamefont {Toschi}}, \
  and\ \bibinfo {author} {\bibfnamefont {J.}~\bibnamefont {Harting}},\ }\href
  {\doibase 10.1002/mats.201100013} {\bibfield  {journal} {\bibinfo  {journal}
  {Macromolecular Theory and Simulations}\ }\textbf {\bibinfo {volume} {20}},\
  \bibinfo {pages} {562} (\bibinfo {year} {2011})}\BibitemShut {NoStop}%
\bibitem [{\citenamefont {Lin}\ and\ \citenamefont {Han}(2002)}]{lin02}%
  \BibitemOpen
  \bibfield  {author} {\bibinfo {author} {\bibfnamefont {A.}~\bibnamefont
  {Lin}}\ and\ \bibinfo {author} {\bibfnamefont {S.-P.}\ \bibnamefont {Han}},\
  }\href@noop {} {\bibfield  {journal} {\bibinfo  {journal} {SIAM Journal on
  Optimization}\ }\textbf {\bibinfo {volume} {13}},\ \bibinfo {pages} {298}
  (\bibinfo {year} {2002})}\BibitemShut {NoStop}%
\bibitem [{\citenamefont {Butler}\ and\ \citenamefont
  {Shaqfeh}(2002)}]{butler02}%
  \BibitemOpen
  \bibfield  {author} {\bibinfo {author} {\bibfnamefont {J.~E.}\ \bibnamefont
  {Butler}}\ and\ \bibinfo {author} {\bibfnamefont {E.~S.~G.}\ \bibnamefont
  {Shaqfeh}},\ }\href@noop {} {\bibfield  {journal} {\bibinfo  {journal} {{J.
  Fluid Mech.}}\ }\textbf {\bibinfo {volume} {468}},\ \bibinfo {pages} {205}
  (\bibinfo {year} {2002})}\BibitemShut {NoStop}%
\bibitem [{\citenamefont {Komnik}\ \emph {et~al.}(2004)\citenamefont {Komnik},
  \citenamefont {Harting},\ and\ \citenamefont {Herrmann}}]{komnik04}%
  \BibitemOpen
  \bibfield  {author} {\bibinfo {author} {\bibfnamefont {A.}~\bibnamefont
  {Komnik}}, \bibinfo {author} {\bibfnamefont {J.}~\bibnamefont {Harting}}, \
  and\ \bibinfo {author} {\bibfnamefont {H.~J.}\ \bibnamefont {Herrmann}},\
  }\href@noop {} {\bibfield  {journal} {\bibinfo  {journal} {Journal of
  Statistical Mechanics: theory and experiment}\ }\textbf {\bibinfo {volume}
  {P12003}} (\bibinfo {year} {2004})}\BibitemShut {NoStop}%
\bibitem [{\citenamefont {Sierou}\ and\ \citenamefont
  {Brady}(2004)}]{sierou04}%
  \BibitemOpen
  \bibfield  {author} {\bibinfo {author} {\bibfnamefont {A.}~\bibnamefont
  {Sierou}}\ and\ \bibinfo {author} {\bibfnamefont {J.~F.}\ \bibnamefont
  {Brady}},\ }\href@noop {} {\bibfield  {journal} {\bibinfo  {journal} {Journal
  of Fluid Mechanics}\ }\textbf {\bibinfo {volume} {506}},\ \bibinfo {pages}
  {285} (\bibinfo {year} {2004})}\BibitemShut {NoStop}%
\bibitem [{\citenamefont {Ball}\ and\ \citenamefont {Melrose}(1995)}]{ball95}%
  \BibitemOpen
  \bibfield  {author} {\bibinfo {author} {\bibfnamefont {R.~C.}\ \bibnamefont
  {Ball}}\ and\ \bibinfo {author} {\bibfnamefont {J.~R.}\ \bibnamefont
  {Melrose}},\ }\href@noop {} {\bibfield  {journal} {\bibinfo  {journal}
  {Advances in colloid and interface science}\ }\textbf {\bibinfo {volume}
  {59}},\ \bibinfo {pages} {19} (\bibinfo {year} {1995})}\BibitemShut {NoStop}%
\bibitem [{\citenamefont {Sierou}\ and\ \citenamefont
  {Brady}(2002)}]{sierou02}%
  \BibitemOpen
  \bibfield  {author} {\bibinfo {author} {\bibfnamefont {A.}~\bibnamefont
  {Sierou}}\ and\ \bibinfo {author} {\bibfnamefont {J.~F.}\ \bibnamefont
  {Brady}},\ }\href@noop {} {\bibfield  {journal} {\bibinfo  {journal} {Journal
  of Rheology}\ }\textbf {\bibinfo {volume} {46}},\ \bibinfo {pages} {1031}
  (\bibinfo {year} {2002})}\BibitemShut {NoStop}%
\bibitem [{\citenamefont {Frijters}\ \emph {et~al.}(2012)\citenamefont
  {Frijters}, \citenamefont {G{\"u}nther},\ and\ \citenamefont
  {Harting}}]{frijters12}%
  \BibitemOpen
  \bibfield  {author} {\bibinfo {author} {\bibfnamefont {S.}~\bibnamefont
  {Frijters}}, \bibinfo {author} {\bibfnamefont {F.}~\bibnamefont
  {G{\"u}nther}}, \ and\ \bibinfo {author} {\bibfnamefont {J.}~\bibnamefont
  {Harting}},\ }\href {\doibase 10.1039/C2SM25209K} {\bibfield  {journal}
  {\bibinfo  {journal} {Soft Matter}\ }\textbf {\bibinfo {volume} {8}},\
  \bibinfo {pages} {6542 } (\bibinfo {year} {2012})}\BibitemShut {NoStop}%
\bibitem [{\citenamefont {Thornton}\ \emph {et~al.}(2013)\citenamefont
  {Thornton}, \citenamefont {Weinhart}, \citenamefont {Ogarko},\ and\
  \citenamefont {Luding}}]{thornton13}%
  \BibitemOpen
  \bibfield  {author} {\bibinfo {author} {\bibfnamefont {A.~R.}\ \bibnamefont
  {Thornton}}, \bibinfo {author} {\bibfnamefont {T.}~\bibnamefont {Weinhart}},
  \bibinfo {author} {\bibfnamefont {V.}~\bibnamefont {Ogarko}}, \ and\ \bibinfo
  {author} {\bibfnamefont {S.}~\bibnamefont {Luding}},\ }\href@noop {}
  {\bibfield  {journal} {\bibinfo  {journal} {Computer methods in materials
  science}\ }\textbf {\bibinfo {volume} {13}},\ \bibinfo {pages} {197}
  (\bibinfo {year} {2013})}\BibitemShut {NoStop}%
\bibitem [{\citenamefont {Jeffrey}\ and\ \citenamefont
  {Onishi}(1984)}]{jeffrey84}%
  \BibitemOpen
  \bibfield  {author} {\bibinfo {author} {\bibfnamefont {D.~J.}\ \bibnamefont
  {Jeffrey}}\ and\ \bibinfo {author} {\bibfnamefont {Y.}~\bibnamefont
  {Onishi}},\ }\href {\doibase 10.1017/S0022112084000355} {\bibfield  {journal}
  {\bibinfo  {journal} {{J. Fluid Mech.}}\ }\textbf {\bibinfo {volume} {139}},\
  \bibinfo {pages} {261} (\bibinfo {year} {1984})}\BibitemShut {NoStop}%
\bibitem [{\citenamefont {Kim}\ and\ \citenamefont {Karrila}(2005)}]{kim05}%
  \BibitemOpen
  \bibfield  {author} {\bibinfo {author} {\bibfnamefont {S.}~\bibnamefont
  {Kim}}\ and\ \bibinfo {author} {\bibfnamefont {S.~J.}\ \bibnamefont
  {Karrila}},\ }\href@noop {} {\emph {\bibinfo {title} {Microhydrodynamics:
  principles and selected applications}}}\ (\bibinfo  {publisher} {Dover
  Publications Inc.},\ \bibinfo {address} {New York},\ \bibinfo {year}
  {2005})\BibitemShut {NoStop}%
\bibitem [{\citenamefont {Clausen}\ \emph {et~al.}(2010)\citenamefont
  {Clausen}, \citenamefont {Reasor~Jr.},\ and\ \citenamefont
  {Aidun}}]{clausen10}%
  \BibitemOpen
  \bibfield  {author} {\bibinfo {author} {\bibfnamefont {J.~R.}\ \bibnamefont
  {Clausen}}, \bibinfo {author} {\bibfnamefont {D.~A.}\ \bibnamefont
  {Reasor~Jr.}}, \ and\ \bibinfo {author} {\bibfnamefont {C.~K.}\ \bibnamefont
  {Aidun}},\ }\href@noop {} {\bibfield  {journal} {\bibinfo  {journal}
  {Computer Physics Communications}\ }\textbf {\bibinfo {volume} {181}},\
  \bibinfo {pages} {1013} (\bibinfo {year} {2010})}\BibitemShut {NoStop}%
\bibitem [{\citenamefont {Lorenz}\ \emph {et~al.}(2009)\citenamefont {Lorenz},
  \citenamefont {Hoekstra},\ and\ \citenamefont {Caiazzo}}]{lorenz09b}%
  \BibitemOpen
  \bibfield  {author} {\bibinfo {author} {\bibfnamefont {E.}~\bibnamefont
  {Lorenz}}, \bibinfo {author} {\bibfnamefont {A.~G.}\ \bibnamefont
  {Hoekstra}}, \ and\ \bibinfo {author} {\bibfnamefont {A.}~\bibnamefont
  {Caiazzo}},\ }\href@noop {} {\bibfield  {journal} {\bibinfo  {journal}
  {{Phys. Rev. E}}\ }\textbf {\bibinfo {volume} {79}},\ \bibinfo {pages}
  {036706} (\bibinfo {year} {2009})}\BibitemShut {NoStop}%
\bibitem [{\citenamefont {{da}~Cunha}\ and\ \citenamefont
  {Hinch}(1996)}]{dacunha96}%
  \BibitemOpen
  \bibfield  {author} {\bibinfo {author} {\bibfnamefont {F.~R.}\ \bibnamefont
  {{da}~Cunha}}\ and\ \bibinfo {author} {\bibfnamefont {E.~J.}\ \bibnamefont
  {Hinch}},\ }\href@noop {} {\bibfield  {journal} {\bibinfo  {journal} {{J.
  Fluid Mech.}}\ }\textbf {\bibinfo {volume} {309}},\ \bibinfo {pages} {211}
  (\bibinfo {year} {1996})}\BibitemShut {NoStop}%
\bibitem [{\citenamefont {Staben}\ \emph {et~al.}(2006)\citenamefont {Staben},
  \citenamefont {Zinchenko},\ and\ \citenamefont {Davis}}]{staben06}%
  \BibitemOpen
  \bibfield  {author} {\bibinfo {author} {\bibfnamefont {M.~E.}\ \bibnamefont
  {Staben}}, \bibinfo {author} {\bibfnamefont {A.~Z.}\ \bibnamefont
  {Zinchenko}}, \ and\ \bibinfo {author} {\bibfnamefont {R.~H.}\ \bibnamefont
  {Davis}},\ }\href@noop {} {\bibfield  {journal} {\bibinfo  {journal} {{J.
  Fluid Mech.}}\ }\textbf {\bibinfo {volume} {553}},\ \bibinfo {pages} {187}
  (\bibinfo {year} {2006})}\BibitemShut {NoStop}%
\end{thebibliography}
\end{document}